\documentclass[12pt,aps,tightenlines,notitlepage,superscriptaddress]{revtex4-1}
\usepackage{amsmath}
\usepackage{amssymb}
\usepackage{bm}
\usepackage[mathscr]{eucal}
\usepackage{theorem}
\usepackage{graphicx}
\usepackage{psfrag}
\usepackage{subfigure}
\usepackage{color}
\usepackage{wasysym}
\include{graphix}
\usepackage{framed}
\usepackage{enumitem}%
\usepackage{lipsum}
\usepackage{float}
\usepackage{stmaryrd}
\usepackage{MnSymbol}
\definecolor{orange}{rgb}{1,0.5,0}

\graphicspath{{figs/}}

\newcommand{\mathd}{\mathrm{d}}
\newcommand{\mathe}{\mathrm{e}}
\newcommand{\imag}{\mathrm{i}}

\newcommand{\vecu}{{\bf{u}}}

\newcommand{\mydrop}{\mathrm{d}p_0/\mathrm{d}x}

\newcommand{\hatA}{\widehat{A}}
\newcommand{\hatB}{\widehat{B}}

\newcommand{\kmax}{\alpha_{\mathrm{max}}}
\newcommand{\cmax}{c_{\mathrm{max}}}
\newcommand{\fmax}{f_{\mathrm{max}}}

\newcommand{\myell}{\ell}
\newcommand{\Ttensor}{\mathbb{T}}
\newcommand{\vecnhat}{\widehat{\bf{n}}}

\begin{document}

%+Title
\title{Linear and Nonlinear Stability Analysis in Microfluidic Systems}
\author{Lennon \'O N\'araigh}
\email{onaraigh@maths.ucd.ie}
\affiliation{School of Mathematics and Statistics, University College Dublin, Belfield, Dublin 4}

\author{Daniel R. Jansen van Vuuren}
\affiliation{School of Mathematics and Statistics, University College Dublin, Belfield, Dublin 4}
\affiliation{School of Engineering, University of Pretoria, cnr Lynnwood Road and Roper Street, Hatfield, Pretoria
South Africa}

\date{\today}

\begin{abstract}
In this article we use analytical and numerical modeling to describe parallel viscous two-phase flows in microchannels.  The focus is on idealized two-dimensional geometries, with a view to validating the various methodologies for future work in three dimensions.  In the first instance, we use analytical Orr--Sommerfeld theory to describe the linear instability which governs the formation of small-amplitude waves in such systems.  We then compare the results of this analysis with an in-house Computational Fluid Dynamics (CFD) solver called TPLS.  Excellent agreement between the theoretical analysis and TPLS is obtained in the regime of small-amplitude waves.  We continue the numerical simulations beyond the point of validity of the Orr--Sommerfeld theory.  In this way, we illustrate the generation of nonlinear interfacial waves and reverse entrainment of one fluid phase into the other.  We justify our simulations further by comparing the numerical results with corresponding results from a commercial CFD code.  This comparison is again extremely favourable -- this rigorous validation paves the way for future work using TPLS or commercial codes to perform extremely detailed three-dimensional simulations of flow in microchannels.
\end{abstract}

\maketitle

\section{Introduction}
\label{sec:intro}

In this work we are concerned with theoretical modelling of interfacial instability of two-phase fluids in microchannels of depth $H\sim 100\,\mu\mathrm{m}$.   Specifically, we are concerned with the instability of the instability of the interface separating the streams of two immiscible liquids.  Such flows are important in microfluidic devices and the related applications in reactions, mixing, emulsions,  and material synthesis~\cite{zhao2011two}.  The intrinsic instability of such flow configurations can be harnessed to promote microfluidic mixing without any active forcing~\cite{HuCubaud2018}.  Previous theoretical works on the subject involve solving the Orr--Sommerfeld equation for the interfacial instability in various parameter regimes.  Such a theoretical approach provides the necessary framework for characterizing the interfacial instability -- at least during the development of the instability, starting from small-amplitude perturbations to the flat interface separating the liquid streams.  However, because the parameter space of the system is large, it is difficult if not impossible to produce a universal characterization of the instability, and previous studies focus on subspaces of the entire parameter space~\cite{naraigh2018instability,Naraigh2014linear}.

Even with the kind of specialization just described, the studies in References~\cite{naraigh2018instability,Naraigh2014linear} are very general.  As such, the aim of this work is to take the methodology of these theoretical works and to apply it to a very particular, detailed, and industrially-relevant test case involving liquid-liquid flow.  This test case is taken from microfluidics, and has been documented in experiments~\cite{HuCubaud2018}.  Therefore, the aim of this paper is to bring theoretical understanding to existing experiments on interfacial flows in microfluidics.

A second aim of this paper is to establish a set of strict benchmarks for the validity of two-phase flow simulations in microchannels.  As such, in this article we are concerned with an idealised two-dimensional system with periodic boundary conditions in the streamwise direction -- apart from a pressure drop driving the flow in the spanwise direction.  No-slip conditions are applied at the channel walls in the other dimension.  In this scenario there is an analytic theory (Linear Stability Analysis, in particular Orr--Sommerfeld Analysis) which predicts exactly what should be the growth rate of a small-amplitude wave on the interface.  We use this theory to establish the correctness of the Computational Fluid Dynamics (CFD) simulations based on an in-house finite-volume two-phase levelset solver; this in turn is used as a base case against which to compare CFD simulations performed with a commercial code.  These simulations then establish the validity of the method which can be used in future work to simulate other microcfluidic flows with more realistic three-dimensioanl geometries.

The layout of the paper is as follows.  In Section~\ref{sec:base} we describe the theoretical model of the base state wherein the two fluid streams are separated by a flat interface.  The system in this case is characterized by the different flow rates in each phase.  We construct an analytical model to predict the width of the two streams (as well as the pressure drop) as a function of the flow rates.  In Section~\ref{sec:method} we present the methodology of the paper, which encompasses both linear stability analysis based on the Orr--Sommerfeld equations, and Direct Numerical Simulations of the fully non-linear two-phase flow equations using a Level-Set Method.  Theoretical results based on this methodology are presented in Sections~\ref{sec:os}--\ref{sec:dns}; Section~\ref{sec:dns} also contains some qualitative comparisons with prior experimental work.  Finally, concluding remarks are presented in Section~\ref{sec:conc}.

\section{Base-State Configuration}
\label{sec:base}

\begin{figure}[htb]
\centering
\includegraphics[width=0.95\textwidth]{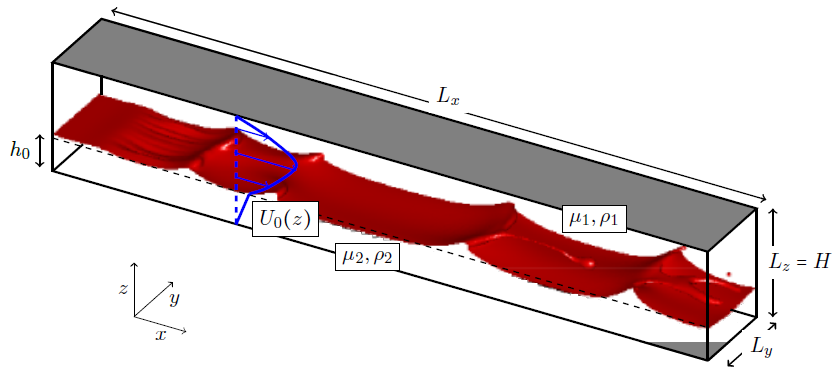}
\caption{Definition sketch showing the problem geometry.  The domain is periodic in the $y$-direction, whereas a pressure drop drives the flow along the $x$-direction.  No-slip conditions are applied at $z=0$and $z=H$.}
\label{fig:sketch1}
\end{figure}

We study the flow of two fluids confined between two parallel plates, shown schematically in Figure~\ref{fig:sketch1}.  
The fluids are assumed to be isothermal and incompressible, with constant densities and viscosities.  The lower layer has density $\rho_1$ and viscosity $\mu_1$.  Correspondingly, the upper layer has density $\rho_2$ and viscosity $\mu_2$.  
The labels $j=1,2$ are used more broadly throughout this work to identify the top ($j=1$) and bottom ($j=2$) fluids. 
The system is assumed to be stably stratified, such that $\rho_2>\rho_1$.
Bounding walls with the implied no-slip boundary conditions are introduced at $z=0$ and $z=H$.  
Periodic boundary conditions are imposed in the spanwise ($y$-) direction (this amounts to a quasi-two-dimensional `parallel plates' configuration).  The boundary conditions in the $x$-direction are left unspecified as yet, save for the imposition of a constant pressure drop $\mydrop$ in that direction.

The system admits an undisturbed or base-state configuration characterized by a Poiseuille flow where the depth of the undisturbed liquid layer is denoted by $h_0$.   In this scenario, the force balance in each phase reads
\begin{equation}
\mu_i\frac{\mathd^2 U_{0i}}{\mathd z^2}=-\mydrop.
\end{equation}
The equations are integrated twice  and the no-slip conditions are applied at $z=0$ and $z=H$.  This yields
\begin{subequations}
\begin{eqnarray}
U_{01}&=&-\frac{1}{2\mu_2}\left|\mydrop\right|(z-H)^2+B(z-H),\\
U_{02}&=&-\frac{1}{2\mu_2}\left|\mydrop\right|z^2+Az,
\end{eqnarray}
\end{subequations}
where $A$ and $B$ are constants of integration.  These are determined by applying continuity of velocity and tangential stress across the interface located at $z=h_0$, hence
\begin{subequations}
\begin{eqnarray}
A&=&\frac{\left[\tfrac{1}{2}\frac{\mu_1}{\mu_2}h_0^2-\tfrac{1}{2}(H-h_0)^2+H(H-h_0)\right]\left|\mydrop\right|}{h_0\mu_1+(H-h_0)\mu_2},\\
B&=&\frac{\left[\tfrac{1}{2}h_0^2-\tfrac{1}{2}\frac{\mu_2}{\mu_1}(H-h_0)^2-Hh_0\right]\left|\mydrop\right|}{h_0\mu_1+(H-h_0)\mu_2}.
\end{eqnarray}%
\label{eq:AB}%
\end{subequations}%
The corresponding flow rates are
\begin{subequations}
\begin{eqnarray}
Q_1&=&H\int_{h_0}^H U_{01}(z)\mathd z,\\
   &=&H\left[-\frac{1}{6\mu_1}\left|\mydrop\right|(H-h_0)^3-\tfrac{1}{2}B(H-h_0)^2\right],\\
Q_2&=&H\int_{0}^{h_0} U_{02}(z)\mathd z,\\
   &=&H\left[-\frac{1}{2\mu_2}\left|\mydrop\right|h^3+\tfrac{1}{2}Ah^2\right].
\end{eqnarray}
\end{subequations}
In what follows, it will be helpful to work with quantities where the dependence on the pressure drop is scaled out.  As such, we introduce $\hatA=A/\left|\mydrop\right|$ and $\hatB=B/\left|\mydrop\right|$.  In this way, the flow rates can be decomposed into a product of a geometric factor, and the pressure drop:
\begin{subequations}
\begin{eqnarray}
Q_1&=&H\left[-\tfrac{1}{6\mu_1}(H-h_0)^3-\tfrac{1}{2}\hatB (H-h_0)^2\right]\left|\mydrop\right|,\\
Q_2&=&H\left[-\frac{1}{6\mu_2}h^3+\tfrac{1}{2}\hatA h^2\right]\left|\mydrop\right|.
\end{eqnarray}
\end{subequations}
The ratio of flow rates is therefore independent of the pressure drop, and given by the formula
\begin{equation}
\varphi=\frac{Q_1}{Q_2}=\frac{-\tfrac{1}{6\mu_1}(H-h_0)^3-\tfrac{1}{2}\hatB (H-h_0)^2}{-\frac{1}{6\mu_2}h^3+\tfrac{1}{2}\hatA h^2}.
\label{eq:varphi}
\end{equation}
In this work we view the flow rates $Q_1$ and $Q_2$ as the key independent variables (along with the channel height $H$).  Thus, the other quantities such as pressure drop and undisturbed interface height $h_0$ emerge as dependent variables, which can be determined via Equations~\eqref{eq:AB}--\eqref{eq:varphi}.  As such, we plot the non-dimensional upper layer depth $\epsilon_1=(H-h_0)/H$ as a function of $\varphi$ for a selected pair of working fluids in Figure~\ref{fig:compare_fig1b_for_paper}.  The selected working fluids are silicon oil ($\mu_2=485\,\mathrm{cP}$, $\rho_2=0.97\,\mathrm{g}\,\mathrm{mL}^{-1}$) and ethanol ($\mu_1=1.08\,\mathrm{cP}$, $\rho_1=0.78\,\mathrm{g}\,\mathrm{mL}^{-1}$).  The surface tension between the two fluids is taken to be $\gamma=1.09\,\mathrm{mN}\,\mathrm{m}^{-1}$ as the surface tension.  The channel height is taken to be $H=250\,\mu\mathrm{m}$. Throughout the paper, we work with these values which are characteristic of two-stream liquid-liquid microfluidic flows, as documented in the experimental work in Reference~\cite{HuCubaud2018}.  
\begin{figure}[htb]
	\centering
		\includegraphics[width=0.6\textwidth]{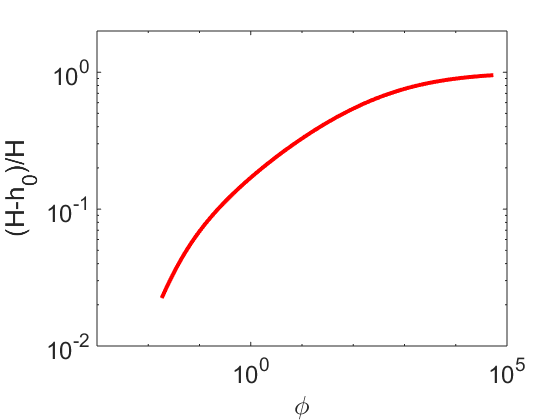}
		\caption{Plot of non-dimensional upper-layer depth $\epsilon_1=(H-h_0)/H$ as a function of the ratio of flow rates $\varphi=Q_1/Q_2$.  The working fluids are the ones given in the main text.}
	\label{fig:compare_fig1b_for_paper}
\end{figure}

It should be emphasized that the experiments in Reference~\cite{HuCubaud2018} are done in a  microchannel geometry, with no-slip boundary conditions in each of the directions normal to the main flow.  This amounts to a fully three-dimensional `square duct' geometry.  In contrast, the base state and detailed linear stability analysis as developed in the present work relies on the existence of periodic boundary conditions in the streamwise direction.  This can be thought of as a quasi-two-dimensional geometry, where the flow is confined between two parallel plates extending to infinity in the $x$- and $y$-directions 
As these two flow profiles are distinct, it is not feasible to compare them quantitatively.  Qualitative comparisons between the two systems are presented  in Section~\ref{sec:dns}, below.

\section{Methodology}
\label{sec:method}

We begin by introducing the governing equations of motion for each phase.  These are the Navier--Stokes equations for viscous incompressible flow:
\begin{subequations}%
\begin{eqnarray}
\rho_i\left(\frac{\partial \vecu_i}{\partial t}+\vecu_i\cdot\nabla\vecu_i\right)&=&-\nabla p_i+\mu_i\nabla^2 \vecu_i
+|dP/dL|\widehat{\bm{e}}_1-\rho_i g \widehat{\bm{e}}_3,\\
\nabla\cdot\vecu_i=0.
\end{eqnarray}%
\label{eq:ns_all}%
\end{subequations}%
where $i=1,2$ labels the phase.  The variable $p_i$ denotes the fluid pressure in the $i^{\text{th}}$ phase.  These equations are solved in a domain similar to that in Figure~\ref{fig:sketch1}: in particular, with no-slip boundary conditions on the channel wall, $\vecu_i(x,y,z=0,t)=\vecu_i(x,y,z=L_z,t)$ and periodic boundary conditions in the $y$-direction, $\vecu_i(x,y,z,t)=\vecu_i(x,y+L_y,z,t)$.  The boundary conditions in the $x$-direction are assumed to be periodic also, with the constant applied pressure drop $|dp/dL|$ driving the flow along the $x$-direction.  The $x$-direction is indicated in Equation~\eqref{eq:ns_all} by the unit vector $\widehat{\bm{e}}_1$.  Finally, the gravitational force is given by $-\rho g\widehat{\bm{e}}_3$, where $g$ is the acceleration due to gravity and $\widehat{\bm{e}}_3$ is the unit vector in the $z$-direction.

The phases $i=1$ and $i=2$ are separated by an interface.  The interface is possibly disconnected and is therefore best described in terms of the zero levelset of a general levelset function, $\phi(x,y,z,t)$.  Hence, the interface is the surface described by the equation $\phi(x,y,z,t)=0$.  The kinematic condition which requires that the interface moves with the flow therefore amounts to the condition that the zero levelset is advected:
\begin{equation}
\frac{\partial\phi}{\partial t}+\vecu\cdot\nabla \phi=0,\qquad \phi=0.
\label{eq:ls}%
\end{equation}
The suppression of the subscript $i$ on the adjective derivative in Equation~\eqref{eq:ls} is deliberate, as it is assumed that the velocity field is continuous across the interface, meaning the distinction between the phases is not necessary there.  The other interfacial conditions involve continuity of tangential stress across the interface, and a jump condition in the normal stress.  Mathematically, these conditions are given as follows:
\begin{equation}
\llbracket \mu_i\vecnhat\cdot\left(\nabla \bm{u}_i+\nabla\bm{u}_i^T\right)\cdot\widehat{\bf{t}}^{(r)}\rrbracket=0,\qquad 
\llbracket \vecnhat\cdot\left[-p_i \mathbb{I}+\mu_i\left(\nabla \bm{u}_i+\nabla\bm{u}_i^T\right)\right]\cdot\vecnhat\rrbracket=\gamma\kappa ,
\label{eq:jump}
\end{equation}%
%
%\label{eq:ns0def_all}%
%\end{subequations}%
%
%
where $\vecnhat$ is a normal vector to the interface (pointing from $i=2$ to $i=1$), and $\widehat{\bf{t}}^{(1)}$ and $\widehat{\bf{t}}^{(2)}$ are the tangent vectors.   The brackets $\llbracket\cdot\rrbracket$ denote the jump condition across the interface ($(i=1)-(i=2)$), and $\kappa$ denotes the interfacial (mean) curvature.  Finally, $\mathbb{I}$ denotes the identity matrix.  It can be emphasized that the base state in Section~\ref{sec:base} represents a flat-interface equilibrium solution of Equations~\eqref{eq:ns_all}--\eqref{eq:jump}.  In the rest of this section we outline the different methodologies that can be used to describe departures from this equilibrium state.

%(pointing from $j=B$ to $j=T$), and $\widehat{\bm{t}}^{(1)}$ and $\widehat{\bm{t}}^{(2)}$ are the tangent vectors.   The brackets $\llbracket\cdot\rrbracket$ denote the jump condition across the interface ($(j=T)-(j=B)$),
%
% B->(2)
% T->(1)

\subsection*{Linear Stability Analysis}

Beyond the equilibrium state described in Section~\ref{sec:base}, we consider the case where a small-amplitude two-dimensional perturbation is introduced around the flat interface, such that the perturbed location of the interface at $t=0$ reads:
\begin{equation}
\phi(x,y,z,\eta)=0,\qquad \eta=h_0+A\sin(\alpha x),\qquad t=0,
\label{eq:eta0}
\end{equation}
where $\alpha$ is the wavenumber of the perturbation and $A$ is the amplitude, with $A\ll h_0$.  The perturbation in Equation~\eqref{eq:eta0}  gives rise to perturbations in the velocity and pressure fields at $t>0$, which in turn feed back into the perturbed interface location such that $\eta$ becomes a function of time, $\eta=\eta(x,t)$. (We use the notation $\delta\vecu_i$ and $\delta p_i$ for the perturbed velocities and pressures, respectively.)  Under the assumption that the initial amplitude $A$ is small, the equations of motion~\eqref{eq:ns_all} can be linearized and the result is a set of evolution equations for the velocity and pressure fluctuations in each phase, as well as the interface location $\eta(x,t)$.  The linearized equations of motion are subject to linearized matching conditions based on Equation~\eqref{eq:jump}.
The solutions of the linearized equations of motion are proportional to an exponential factor of the form $\mathe^{\lambda t+\imag \alpha x}$.  Substituting this trial solution into the equations of motion gives an eigenvalue problem for the eigenvalues $\lambda$.  The eigenvalues depend on $\alpha$; the functional form of this dependence is called the dispersion relation.

For a given $\alpha$, we compute the eigenvalue with the largest real part, denoted here by $\lambda_0(\alpha)$.  The real part of $\lambda_0(\alpha)$ is plotted as a function of $\alpha$ -- this is denoted by $\lambda_{0\mathrm{r}}(\alpha)$.  If $\lambda_{\mathrm{r}0}(\alpha)>0$ for some $\alpha$, then the base state is linearly unstable -- for a particular wavenumber the initial small-amplitude disturbance is thereby guaranteed to grow exponentially.  On the other hand, if $\lambda_{\mathrm{r}0}(\alpha)<0$ the base state is linearly stable and the disturbance is guaranteed to die out as $t\rightarrow \infty$.  The crossover between these two scenarios occurs when $\lambda_{\mathrm{r}0}(\alpha)\leq 0$, and $\lambda_{\mathrm{r}0}(\alpha)=0$ for a discrete number of values of $\alpha$ -- this is called criticality.

In practice, it is straightforward (if tedious) to derive the linearized equations of motion and from there, to pass over to the eigenvalue analysis and hence, to compute the dispersion relation numerically.  The technique for doing this is described in Appendix~\ref{app:lsa}.
The resulting equations constitute an eigenvalue problem for the  streamfunction components $(\Psi_2,\Psi_1)$ defined in Appendix~\ref{app:lsa}, with eigenvalue $\lambda$.  These equations can be formulated in an operator/matrix form given and hence, solved  numerically using standard Chebyshev collocation techniques.  This numerical method  is now well established~\citep{Boomkamp1997,Sahu07,naraigh2018instability,Naraigh2014linear} and is used without further commentary in this work.

\subsection*{Computational Fluid Dynamics -- TPLS}

Beyond linear theory, numerical simulation is required to describe the interfacial dynamics of the two-phase flow.  We introduce TPLS -- an in-house Computational Fluid Dyanmics solver based on the levelset method with a continuous surface tension model~\citep{Sussman1999}.  Such a levelset method can be viewed as a very realstic approximation to the two-phase Navier--Stokes equations with the sharp interfacial conditions~\eqref{eq:jump}.  However, the levelset method represents a great simplification, as it is essentially a one-fluid model: the two phases are treated as one continuous fluid, and the material properties (density, viscosity) transition smoothly from one set of values to another across a narrow width $\epsilon$.  A levelset function $\phi$ is used to track the interface location: the interface is specified by the value $\phi=0$; otherwise, $\phi$ is given by the signed distance to the interface; the sign of $\phi$ is therefore used to label unambiguously the two fluid phases.  In the same manner, the force due to the surface tension is redistributed across a volume the same thickness $\epsilon$, centred around the interface.     Hence, one obtains a single momentum equation, valid throughout the entire flow domain:
\begin{subequations}
\begin{equation}
\rho(\phi)\left(\frac{\partial \vecu}{\partial t}+\vecu\cdot\nabla\vecu\right)=-\nabla p+\nabla\cdot\left[\mu\left(\nabla \vecu+\nabla\vecu^T\right)\right]
+\gamma\delta_\epsilon(\phi)\kappa \vecnhat 
+|dp/dL|\widehat{\bm{e}}_1
-\rho(\phi)g\widehat{\bm{e}}_3,
%-\frac{1}{\mywe}\delta_\epsilon(\phi)\vecnhat\nabla\cdot\vecnhat-\mygrav\rho(\phi)\veczhat,
%-\rho \grav \mathbf{e}_z,
\end{equation}
\begin{equation}
\nabla\cdot\vecu=0.
\end{equation}%
\label{eq:ls_mom}%
\end{subequations}%
%
%\rho&=&r\left(1-H(\phi)\right)+H(\phi),\\
%\mu&=&m\left(1-H(\phi)\right)+H(\phi),\\
%
%
%
%
%
The quantities $\kappa$ and $\vecnhat$ in Equation~\eqref{eq:ls_mom} are geometric objects and correspond to the mean interfacial curvature and interfacial unit normal vector respectively.  These can be obtained from derivatives of the levelset function $\phi$ as follows:
\begin{equation}
\vecnhat=\frac{\nabla\phi}{|\nabla\phi|},\qquad \kappa=-\nabla\cdot\vecnhat,\qquad \frac{\partial \phi}{\partial t}+\vecu\cdot\nabla\phi=0,
\label{eq:ls_def}
\end{equation}%
Equally, the levelset function determines how the surface-tension force is distributed over a small volume of width $\epsilon$ centred at the interface, via the relation
\begin{equation}
\delta_\epsilon(\phi)=\frac{\mathd H_\epsilon}{\mathd \phi}.
\end{equation}
Here, $H_\epsilon(\phi)$ is a smoothened step function, such that $H_\epsilon(\phi)=0$ as $\phi\rightarrow -\infty$, $H_\epsilon(\phi)=1$ as $\phi\rightarrow \infty$, and such that $H_\epsilon$ transitions smoothly from $H_\epsilon=0$ to $H_\epsilon=1$ across a narrow gap of thickness $\epsilon$.  In this way, the one-fluid density and viscosity can also be introduced, and transition smoothly from one set of constant values to another, corresponding to the different fluid phases:
\begin{subequations}
\begin{eqnarray}
\rho(\phi)&=&\rho_2\left(1-H_\epsilon(\phi)\right)+\rho_1H_\epsilon(\phi),\\
\mu(\phi)&=&\mu_2\left(1-H_\epsilon(\phi)\right)+\mu_1H_\epsilon(\phi).
\end{eqnarray}
\label{eq:rhodef}
\end{subequations}
Thus, we are adopting a convention where $\phi<0$ in the bottom layer (phase 2) and $\phi>0$ in the top layer (phase 1).

Equations~\eqref{eq:ls_mom}--\eqref{eq:rhodef} are solved in a density-contrast implementation of the computational framework TPLS~\citep{Naraigh2014linear,tpls_code}.   Specifically, the equations are discretized in space using an isotropic MAC grid wherein vector quantities are defined at cell faces and scalar quantities at the respective cell centres.  In terms of the temporal discretization, a third-order Adams--Bashforth scheme is used to treat the convective derivative, while the momentum fluxes are treated using the Crank--Nicolson method.  Pressure and the associated incompressibility of the flow are treated using a standard projection  method~\citep{chorin1968numerical}.  This computational methodology concerning the basic hydrodynamics is explained in an expository fashion in \cite{fannon2015}.
The levelset function is advected using a third-order (fifth-order accurate) WENO scheme~\citep{jiang1996efficient}, which  is subsequently reinitialized using a Hamilton--Jacobi equation. 
%
%
% We have experimented with several reinitialization methods, including the algorithms of Russo and Smereka~\cite{russo2000}, Hartmann~\cite{hartmann2008differential}, and Sussman and Fatemi~\cite{sussman1999efficient}.  \commentpdms{So which is used here then?} For the simulations presented herein all methods gave near-identical results \commentpdms{update this, not necessarily true at low $\mywe$.}, albeit that the mass loss was minimized using the method of \cite{Sussman1999}.
%
%
%
Validation tests of the method can be found in prior works~\cite{solomenko2016,Naraigh2014linear} -- in particular, the code reproduces all the results of linear theory.  A sample grid-refinement study pertinent to the present work is presented in Appendix~\ref{app:cvg}, herein.  Finally, it can be noted that the numerical simulations of Equations~\eqref{eq:ls_mom}--\eqref{eq:rhodef} are implemented in a Fortran 90 code (TPLS), using a domain decomposition with a maximum of 40 MPI processes, on a machine with a $2\times 20$ core 2.4 GHz Intel Xeon Gold CPU.

\subsection*{Computational Fluid Dynamics -- ANSYS Fluent}

One of the main aims of this paper is to establish a set of strict benchmarks for the validity of two-phase flow simulations in microchannels.   As such, we compare simulations of the same underling system using different methodologies.  For this reason, we further perform simulations of the channel flow using ANSYS Fluent 19 -- in addition to the TPLS simulations just described.  The simulation setup in ANSYS Fluent involves the laminar model.  The pressure-velocity coupling is done using the SIMPLE algorithm, together with the PREssure STaggering Option (PRESTO!) for the pressure solver.  A Second-Order Upwind scheme is used for the interpolation of momentum, with a First-Order Upwind scheme for the other parameters.  The time-stepping is implicit.
For the interface capturing,  we use the Volume of Fluid (VOF) method with coupled level set to capture the interfacial matching conditions and properly resolve the interface.    
The simulations are carried out on a machine with a single Intel i5-6200U processor (2 cores, 4 threads, 2.3 GHz).

\subsection*{Justification of focus on two-dimensional systems}

Throughout this work, we focus on strictly two-dimensional disturbances to the base flow, both in the linear regime where linear stability analysis and Orr--Sommerfeld theory are valid, and also, in the nonlinear regime where the methods of Computational Fluid Mechanics (CFD) are most pertinent.    Certainly, there are pragmatic reasons for this, as the resulting two-dimensional studies (both in linear theory and in the CFD) are easier to perform than would be the case in three dimensions.  At the same time, there are solid theoretical reasons for this.  Certainly, no Squire's theorem does not exist for three dimensions~\citep{Yiantsios1988}, meaning it is not true \textit{a priori} that the most-amplified wave in linear theory is two-dimensional.  However, in practice, the most-amplified wave in  linear theory is usually two-dimensional~\citep{sahu2011three}.  Furthermore, many of the results from two-dimensional studies  carry over to three dimensions.  For instance, the formation of large-amplitude waves in a two-dimensional problem tends to imply the formation of similar waves in the corresponding three-dimensional problem~\cite{naraigh2018instability}.  Hence, we are justified in this work in focusing only on two-dimensional systems. 

A further justification is that in this article we are concerned with establishing simplified test cases against which CFD codes can be validated for the purpose of performing two-phase microfluidic simulations.  Since we focus in the first instance on test cases in which an analytical comparator is available (Orr--Sommerfeld theory), this provides a strict benchmark for the accuracy of the CFD simulations, and a knowledge base to pursue more complicated CFD calculations in future work (in particular, using realistic three-dimensional geometries).

\section{Results -- Linear Stability Analysis}
\label{sec:os}

In this section we perform a linear stability analysis for a range of flow rates, for the fixed working fluids given in Section~\ref{sec:base} (silicone oil and ethanol), with the channel geometry fixed also (specifically, $H=250\,\mu\,\mathrm{m}$).  The aim of this section is to characterize the linear stability of the quasi-two-dimensional parallel-plates geometry for a range of flowrates typical of flow in microchannels.

\begin{figure}
	\centering
		\includegraphics[width=0.8\textwidth]{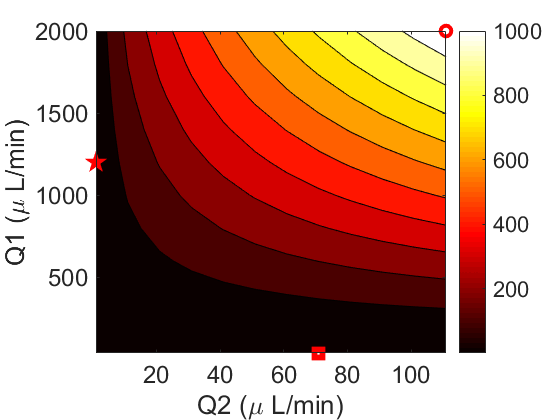}
		\caption{Plot of the growth rate of the most-dangerous mode as a function of flow-rates.  The growth rates are measured in $\mathrm{Hz}$.  The points marked by the square, the star, and the circle are singled out for in-depth study in what follows.}
	\label{fig:growth_rate_scan}
\end{figure}

Accordingly, a linear stability analysis is performed for different pairs of flow rates $(Q_1,Q_2)$ characteristic of microchannels, as specified in Reference~\cite{HuCubaud2018}.  For each flowrate pair, the growth rate $\lambda(\alpha)=\lambda_{\mathrm{r}}(\alpha)+\imag \lambda_{\mathrm{i}}(\alpha)$ is computed for a range of different wavenumbers $\alpha$.  The most-dangerous mode is selected -- this is the wavenumber $\alpha$ that that maximizes $\lambda_{\mathrm{r}}(\alpha)$ (denoted hereafter by $\kmax$).  The growth rate of the most-dangerous mode is then plotted in Figure~\ref{fig:growth_rate_scan}.
The growth rate is positive for each flowrate pair, confirming the system is linearly unstable in the considered operational range.

\begin{figure}
	\centering
		\includegraphics[width=0.8\textwidth]{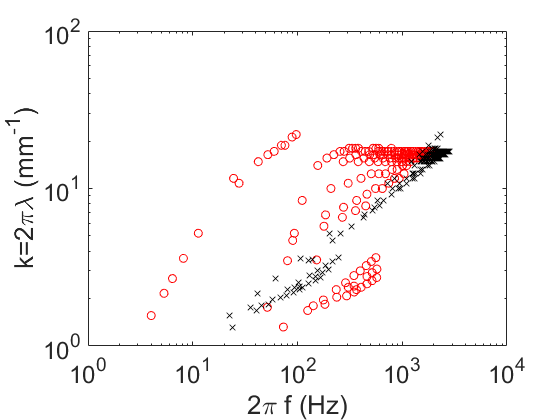}
		\caption{Dispersion relationship between the most-dangerous mode $\kmax$ and the corresponding wave frequency $2\pi \fmax$, for the entire range of considered flow rates.  Circles: Linear Stability Theory and Orr--Sommerfeld Analysis.  Crosses: Inviscid theory and Equation~\eqref{eq:inviscid}.}
	\label{fig:k_vs_f}
\end{figure}

We attempt to classify the observed instabilities, starting in the first instance with a simplified approach.  As such, we first of all look at the most-dangerous mode for each considered flowrate.  Specifically, we look at $\kmax$, the corresponding wave speed $\cmax=-[\lambda_{\mathrm{i}}(\kmax)]/\kmax$, and the corresponding frequency $\fmax=\alpha \cmax/(2\pi)$.  We determine to what extent the wave frequency can be predicted using the formulae for gravity-capillary waves in stratified inviscid two-phase flows~\cite{Acheson1990},
\begin{equation}
c^{\text{gp}}=V_i+ \bigg\{\left[\left(\frac{\rho_2-\rho_1}{\rho_2+\rho_1}\right)\frac{g}{\alpha}+\frac{\alpha \gamma}{\rho_1+\rho_2}\right]\tanh[\alpha (h_0/H)] \bigg\}^{1/2},\qquad
f^{\text{gp}}=\alpha c^{\text{gp}}/(2\pi).
\label{eq:inviscid}
\end{equation} 
Here, $V_i$ denotes the velocity of the interface in the laboratory frame.  As such, in Figure~\ref{fig:k_vs_f} we re-plot the data from Figure~\ref{fig:growth_rate_scan} in a new form: for the most-dangerous mode of each flow-rate pair, we plot $2\pi \fmax$ on the horizontal axis and $\kmax$ on the vertical axis, to build up a comprehensive scatter plot.  We compare the results with notional values from the inviscid formula in Equation~\eqref{eq:inviscid}.
From the figure, it can be seen that the dispersion relation ($\alpha$ versus $2\pi f$) for the linear stability analysis shows a considerable spread in values compared to the inviscid theory in Equation~\eqref{eq:inviscid}.  Therefore, it can be concluded that the inviscid formula is not adequate as a means of classifying the instability in the different parts of the flow-pattern map in Figure~\ref{fig:growth_rate_scan}.  A different method of classifying the instability is therefore required.

\begin{figure}
	\centering
		\subfigure[]{\includegraphics[width=0.45\textwidth]{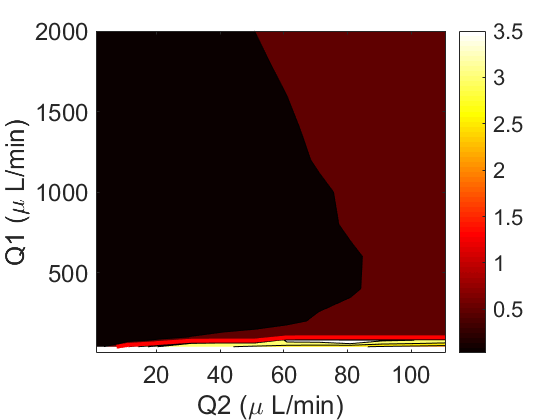}}
		\subfigure[]{\includegraphics[width=0.45\textwidth]{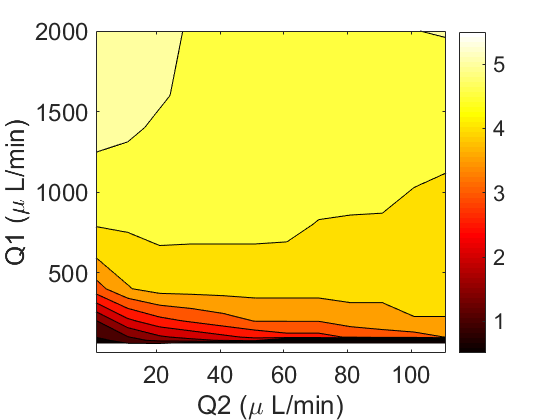}}
		\subfigure[]{\includegraphics[width=0.45\textwidth]{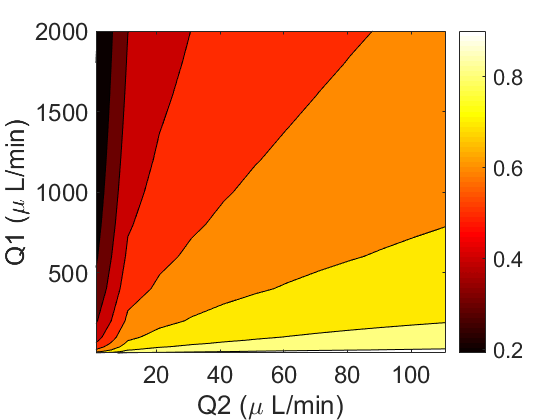}}
		\caption{A more in-depth view of the stability analysis, with key variables plotted as a function of flow rates $Q_2$ and $Q_1$.  (a) $\cmax/c^{\text{gp}}$; (b) $\kmax H$; (c) $h_0/H$.  In all cases the indicated scale on the colorbar is dimensionless.}
	\label{fig:scans_many}
\end{figure}

As such, in Figure~\ref{fig:scans_many} we present:
\begin{enumerate}[noitemsep,label=\alph*]
\item A plot of $\cmax/c^{\text{gp}}$, i.e. the wavespeed of the most-dangerous mode as a function of flow rates, normalized by the notional inviscid wave;
\item A plot of the $\kmax H$ as a function of flow rates;
\item A plot of the film depth as a function of flow rates.
\end{enumerate}
Plots~\ref{fig:scans_many}(a) and (b) depend on the Orr--Sommerfeld analysis, whereas plot (c) depends on the base state only.  From (a) the spread in wave speeds with respect to the inviscid theory can again be observed.  However, the plot gives further information, as it enables one to distinguish between slow and vast waves: the fast waves are confined to a narrow horizontal band in the south of the flow-pattern map and are characterized by:
\begin{itemize}[noitemsep]
\item {\textbf{Fast waves}}: high speed $\cmax/c^{\text{gp}}>1$;
\item Long wavelengths $\alpha H\apprle 1$, hence $\lambda \approx 2\pi H$;
\item Low top-layer flow rates.
\end{itemize}
In contrast, the slow waves are found throughout the rest of the flow-pattern map and are characterized by:
\begin{itemize}[noitemsep]
\item {\textbf{Slow waves}}: low speed $\cmax/c^{\text{gp}}<1$;
\item Shorter wavelengths $\alpha H\approx 5$, hence $\lambda \approx H$;
%\item Shallow bottom layers $h_0/H\apprge 0.2$;
\item High top-layer flow rates.
\end{itemize}

\begin{figure}
	\centering
		%\subfigure[Case A -- Circle]{\includegraphics[width=0.45\textwidth]{dispersion_circle}}
		%\subfigure[Case B -- Square]{\includegraphics[width=0.45\textwidth]{dispersion_square}}
		%\subfigure[Case C -- Pentagram]{\includegraphics[width=0.45\textwidth]{dispersion_pent}}
		%\subfigure[Case D -- Diamond]{\includegraphics[width=0.45\textwidth]{dispersion_diamond}}
		\subfigure[\,\,Sq]{\includegraphics[width=0.45\textwidth]{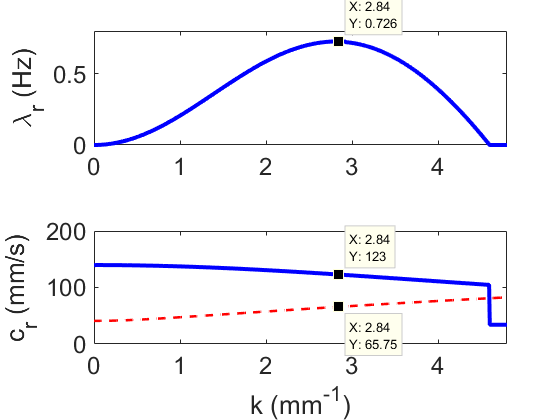}}
		\subfigure[\,\,St]{\includegraphics[width=0.45\textwidth]{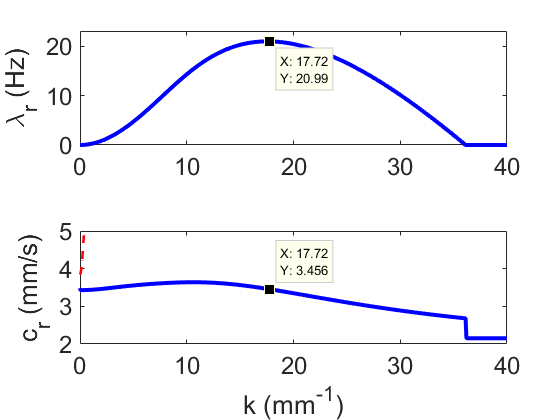}}
		\subfigure[\,\,Circ]{\includegraphics[width=0.45\textwidth]{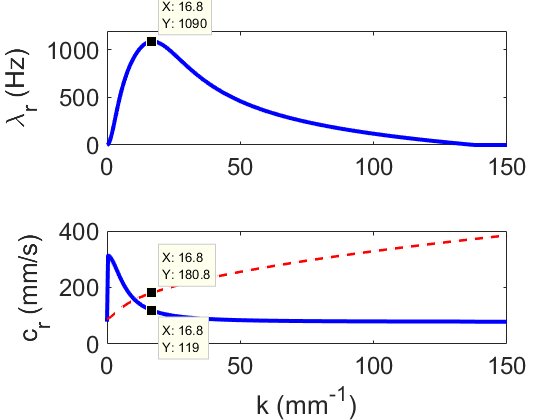}}
\caption{Growth rate and wave speed for the various special cases.  The growth rate is shown at the top in each panel and the wave speed is shown at the bottom.  For wave speeds, a comparison is made with the theory of inviscid gravity-capillary waves (broken line).}  
	\label{fig:disp_special}
\end{figure}

In order to understand this classification in more depth, we carry out a more in-depth linear stability analysis for three representative test cases {\textbf{St}} (Marked with the star in Figure~\ref{fig:growth_rate_scan}), {\textbf{Sq}}  (Marked with the square in Figure~\ref{fig:growth_rate_scan}), and {\textbf{Circ}}  (Marked with the circle in Figure~\ref{fig:growth_rate_scan}).
The properties of these cases are given in detail in Table~\ref{tab:special}.
\begin{table}
	\centering
		\begin{tabular}{|c|c|c|c|c|c|c|c|}
		\hline
			Case & %1
			Symbol &  %2
			$Q_1$ ($\mu\,\mathrm{L}/\mathrm{min}$)  & %3
			$Q_2$ ($\mu\,\mathrm{L}/\mathrm{min}$)  & %4
			$c_{\mathrm{r}}(\kmax)/c^{\mathrm{gp}}(\kmax)$ & %5
			$c_{\mathrm{r}}(\kmax)/V_i$ & %6
			$\kmax$ ($\mathrm{mm}^{-1}$) &  % 7 
			$\lambda_{\mathrm{r}}(\kmax)$ ($\mathrm{Hz}$)  \\ %8
			\hline
			\hline
			%   1            xx  2         xx 3    xx  4  xx  5     xx     6   xx   7   xx   8
			{\textbf{Sq}}    & Square     &  40   &   71 & 2.65   &   3.56   &  2.84 &  0.726  \\
      {\textbf{St}}    & Star       &  1200 &    1 & 0.0385 &   1.42   & 17.72 &  20.99  \\
			{\textbf{Circ}}  & Circle     &  2000 &  110 & 0.708  &   1.52   & 16.8  &  1089  \\
			\hline 
		\end{tabular}
\caption{Special cases chosen for in-depth study.  The cases correspond to the highlighted datapoints (square, star, and circl) in Figure~\ref{fig:growth_rate_scan}.}
\label{tab:special}
\end{table}
The full dispersion relation $\lambda(\alpha)$ is shown for each test case in  Figure~\ref{fig:disp_special}.  
As these special cases are motivated by the scans through the parameter case in Figure~\ref{fig:scans_many}, they exemplify fast waves ({\textbf{Sq}}) and slow waves ({\textbf{St}}, {\textbf{Circ}}).  Cases ({\textbf{Sq}}) and ({\textbf{St}}) are diametric opposites: the most-dangerous mode occurs at a relatively long wavenumber in case Sq: specifically, $\kmax=2.84\,\text{mm}^{-1}$, hence $\lambda/H\approx 2\pi/0.71$, hence, a wavelength much greater than the channel height.  
In case ({\textbf{St}}), $\kmax \approx 20\,\text{mm}^{-1}$, hence $\lambda/H\approx 2\pi/5$, i.e. a wavelength comparable to the channel height.  Case ({\textbf{Circ}}) also corresponds to a slow wave, but can be viewed more as an interemediate case between the extremes ({\textbf{Sq}}) and ({\textbf{St}}).  Specifically, $\kmax=16.8\,\text{mm}^{-1}$, hence $\lambda/H\approx 2\pi/4.2$,
hence $\lambda \approx 1.5 H$.  The real reason for including the intermediate case ({\textbf{Circ}}) can be seen from the last column of the table -- quite clearly, it is the case with the largest growth rate $\lambda_{\mathrm{r}}(\alpha_{\mathrm{max}})$.

In each case in Figure~\ref{fig:disp_special} the wave speed $c_\mathrm{r}$ is analysed -- this is computed from the eigenvalue analysis via $c_\mathrm{r}=-[\lambda_{\mathrm{i}}(\alpha)]/\alpha$.  For comparison, the wavespeed of a gravity-capillary wave is again also shown. It can be noted that for large density ratios, good agreement between Equation~\eqref{eq:inviscid} and the full eigenvalue analysis has been established~\cite{naraigh2018instability}.  However, as we are working with a small density ratio (specifically, $r=0.97/0.78$), the lack of agreement between the two theories is not surprising.

\begin{figure}
	\centering
		\subfigure[\,\,Sq]{\includegraphics[width=0.45\textwidth]{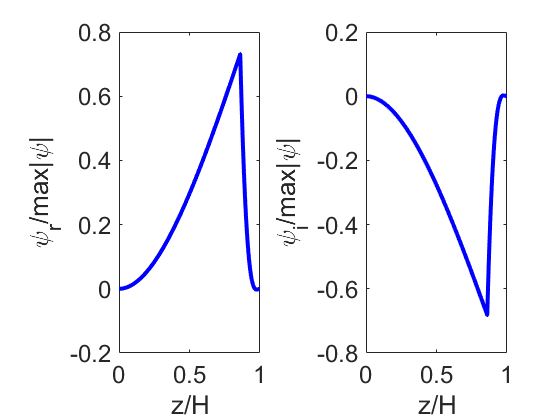}}
		\subfigure[\,\,St]{\includegraphics[width=0.45\textwidth]{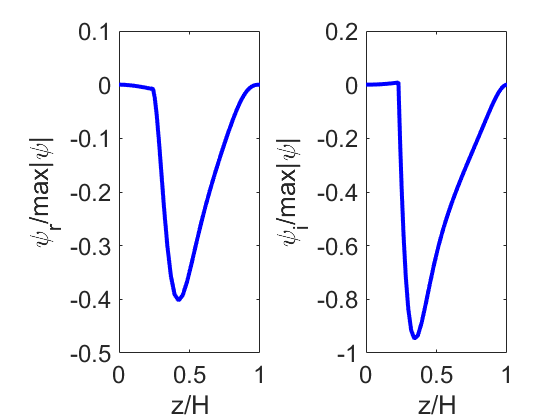}}
		\subfigure[\,\,Circ]{\includegraphics[width=0.45\textwidth]{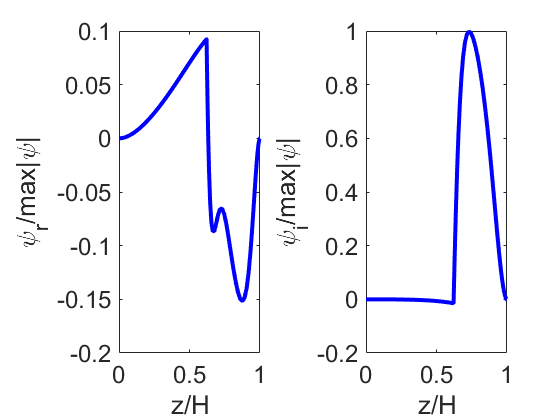}}
		\caption{Streamfunction evaluated at the most-dangerous mode for the various special cases.}
	\label{fig:sf_special}
\end{figure}

The special cases are looked at in from another point of view in Figure~\ref{fig:sf_special} where the streamfunction of the disturbance at the most-dangerous mode is plotted.
Each special case involves a streamfunction that is sharply peaked at the interface, suggesting that the instability is interfacial in nature. 
Case ({\textbf{Sq}}) involves a streamfunction whose largest non-zero component is in the lower layer; Case ({\textbf{St}}) is the opposite.  Case ({\textbf{Circ}}) is intermediate between ({\textbf{Sq}}) and ({\textbf{St}}), with a significant non-zero component in both layers.
 
Using the functional form of these streamfunctions,  we have also performed an  energy-budget analysis, whereby the energy-balance equation
\[
\mathcal{P}=\frac{\mathd }{\mathd t}\int \tfrac{1}{2}\rho_1 |\delta \vecu_1|^2\mathd ^2 x+
\frac{\mathd }{\mathd t}\int \tfrac{1}{2}\rho_2 |\delta \vecu_2|^2\mathd ^2 x
\]
is analysed using the linearized equations of motion.  Here $\delta\vecu_1$ and $\delta\vecu_2$ denote the disturbance velocities, i.e. the velocity over and above the base-state value; these can be obtained from the streamfunctions in Figure~\ref{fig:sf_special}.  The power $\mathcal{P}$ is decomposed into different parts, and it is found that the main positive contribution to $\mathcal{P}$ is due to the so-called interfacial term, which derives from the viscosity mismatch across the interface. As such, the instability is a classical Yih instability~\cite{Yih1967}.  These terms are made precise in Appendix~\ref{app:lsa}.

\section{Results -- Computational Fluid Dynamics}
\label{sec:dns}

To further illustrate the behaviour of the interfacial waves in linear theory -- and to illustrate the behaviour of the waves beyond the linear theory, in this section we carry out numerical simulations for the test case {\textbf{Circ}}, using the TPLS in-house solver.  The numerical parameters for the simulations are: $N_x=440$ gridpoints in the $x$-direction, $N_z=330$ gridpoints in the $z$-direction,  $\Delta t=10^{-5}$; here, $\Delta t$ is the timestep.  In this way, the longest wave that fits inside the computational domain corresponds to the wavelength of the most-dangerous mode of the linear theory.  
These parameters are sufficient for the numerical simulations to demonstrate grid-independence -- see Appendix~\ref{app:cvg} for full justification.  The numerical simulations are performed in non-dimensional variables, such that the rescaled channel height in is unity, and the scaled time is $\tau=tU_p/H$, where $U_p$ is the friction velocity given by $U_p=\sqrt{(H/\rho_1)|dP/dL|}$.

The numerical simulations are seeded with a zero initial velocity field and a zero initial pressure field.  The perturbation with respect to the equilibrium solution is provided by way of an initial wavy interface profile:
\begin{equation}
\eta(x,z,t=0)=h_0+\epsilon\sum_{j=1}^N \cos(j(2\pi/L_x)x+\varphi_j),
\label{eq:ic}
\end{equation}
where $\varphi_j$ is a random phase and $\epsilon$ is an amplitude.  The values $\epsilon/H=5\times 10^{-3}$ and $N=5$ are chosen.
For the considered test case (i.e. {\textbf{Circ}}), the maximum dimensionless growth rate obtained from the linear theory is computed to be $g=\lambda_{\mathrm{r}}(\alpha_{\mathrm{max}})H/U_p=0.3422$.  In this way, the initial amplitude $\epsilon$ is amplified as time goes by, $\epsilon\rightarrow \epsilon\times (\mathe^{g\tau})$, where $\tau=tH/U_p$ is dimensionless time.  As such,  in order for significant wave growth to be observed (defined as $\eta(x,z,t)/H\geq 0.1$), the simulation must run for a dimensionless time of at least $\tau=(1/g)\ln(0.1/0.005)=8.75$.  Instead, have run the simulations out to $\tau=10$, which requires $48$  hours on the machine described in Section~\ref{sec:method}.   We notice in passing that the dimensionless growth rates for the other test cases ({\textbf{Sq}} and {\textbf{St}}) have $g\leq 0.02$.  For significant wave growth to occur in  these test cases, the simulation time would have to be extended by a factor of at least $0.342/0.02\approx 17.1$, which is computationally infeasible with the hardware described in Section~\ref{sec:method}.  In any case, the numerical simulations of the test case {\textbf{Circ}} are detailed enough to suffice for the purpose of understanding the evolution of the waves beyond the linear theory.

\begin{figure}
	\centering
		\subfigure[]{\includegraphics[width=0.45\textwidth]{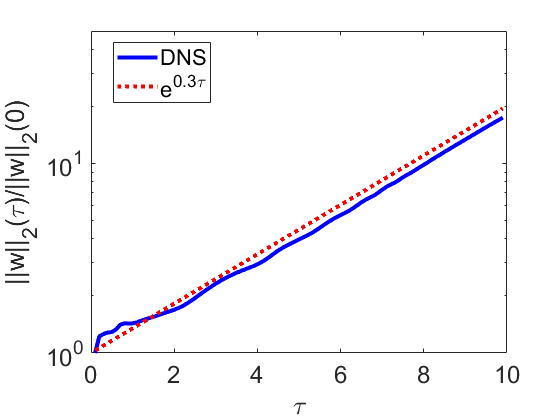}}
		\subfigure[]{\includegraphics[width=0.45\textwidth]{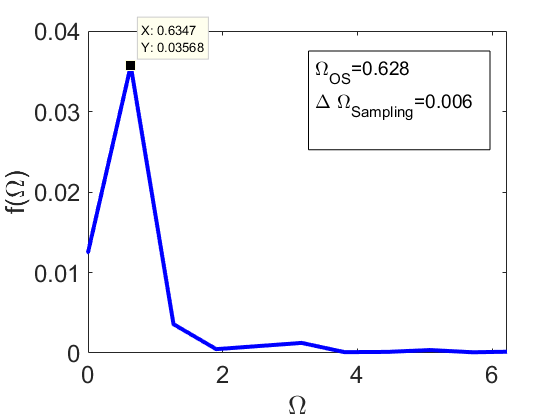}}
		\caption{Comparison between the results of the Direct Numerical Simulations and linear theory.}
	\label{fig:validate_lsa}
\end{figure}

We first of all present results for the growth of the small-amplitude wave~\eqref{eq:ic}.  In Figure~\ref{fig:validate_lsa}(a) we plot the $L^2$ norm of the wall-normal velocity, 
\begin{equation}
\|w\|_2(\tau)=\sqrt{\iint w^2(x,z,\tau)\mathd x\,\mathd z},\qquad \tau=t(U_p/H)
\label{eq:l2_norm}
\end{equation}
as a function of the dimensionless time $\tau$.
From the figure, it can be seen that $\|w\|_2(t)$ grows exponentially, with $\|w\|_2(t)/\|w\|_2(0)\approx \mathe^{0.3 t}$, close to the theoretical maximum rate $g=0.3422$.  The result whereby the observed numerical growth rate is less than the maximum growth rate is due to the fact that $\|w\|_2(t)$ in Equation~\eqref{eq:l2_norm} contains a mixture of different modes (i.e. the most dangerous mode and overtones, as indicated by the initial condition~\eqref{eq:ic}).  To illustrate this further, we have computed the power-spectral density
\[
f(\Omega)=|\text{F.T.}[w(x=L_x/2,h_0,\tau)]|^2,\qquad \Omega=\omega (H/U_p),
\]
where $\text{F.T.}$ stands for Fourier transform, taken in the time-frequency domain, and where $\Omega$ denotes the dimensionless value of the frequency ($\omega$ denotes the corresponding dimensional value of the frequency).
  The power spectral density is plotted in Figure~\ref{fig:validate_lsa}(b).  There, there is a large maximum at $\Omega=0.628$.  This is very close to the most-dangerous mode of linear theory, $\Omega= 0.635$.  The discrepancy between the two values can be explained by the sampling frequency of the numerical simulations: the quantity $w(x=L_x/2,h_0,\tau)$ is sampled at a rate $\Delta \tau=0.1$ (dimensionless time units), the maximum in the $f(\Omega)$ corresponds to the maximum frequency in the numerical solutions to within a tolerance $\Delta \Omega=\Omega^2 \Delta \tau/(2\pi)$, hence $\delta \Omega=0.006$.  In this way, the observed maximum frequency in the simulation is consistent with the assumption that this frequency is obtained from the most-dangerous mode of the linear theory.

\begin{figure}
	\centering
		\subfigure[$\,\,\tau=15$]{\includegraphics[width=0.45\textwidth]{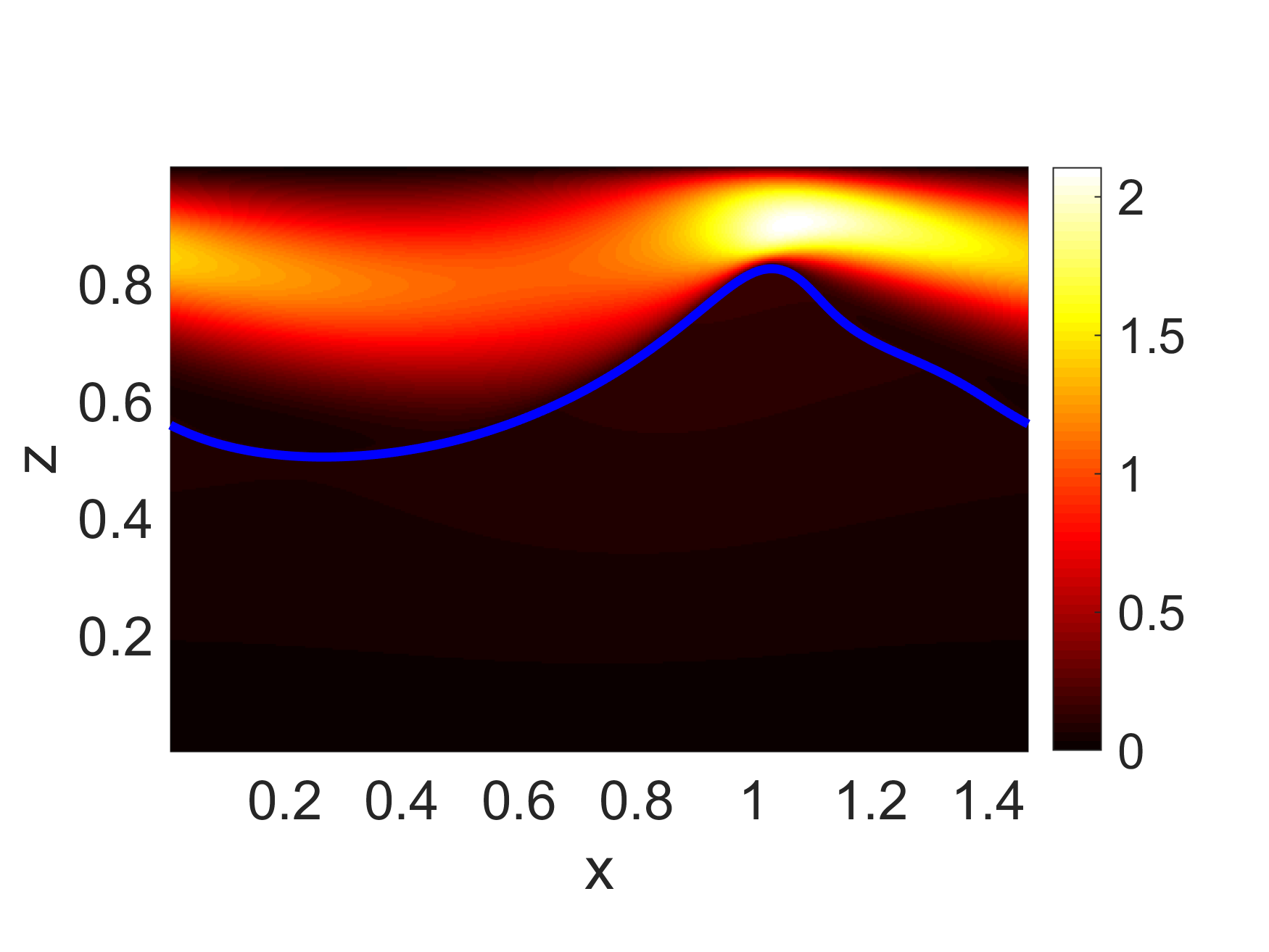}}
		\subfigure[$\,\,\tau=16.5$]{\includegraphics[width=0.45\textwidth]{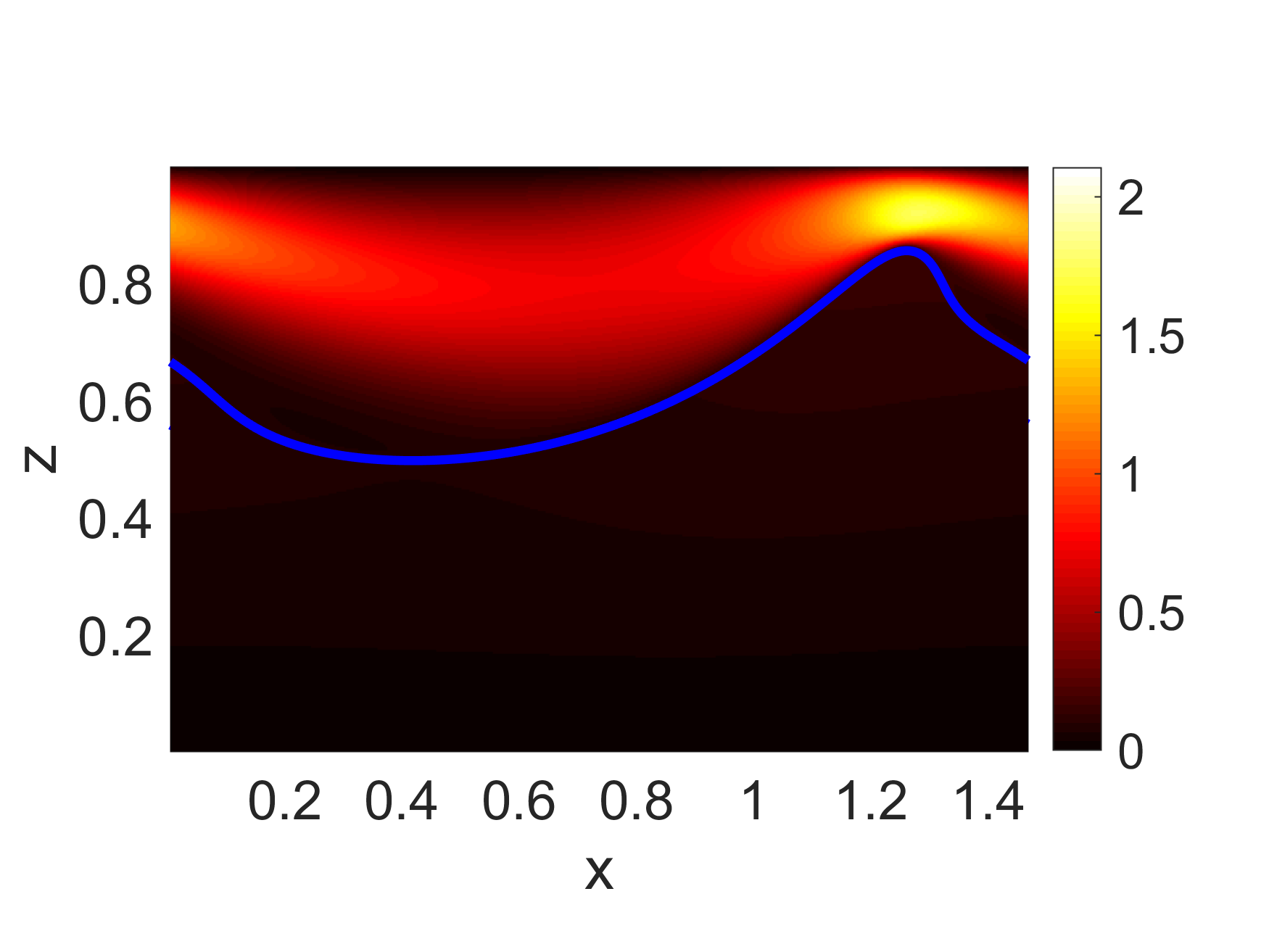}}
		\subfigure[$\,\,\tau=20.0$]{\includegraphics[width=0.45\textwidth]{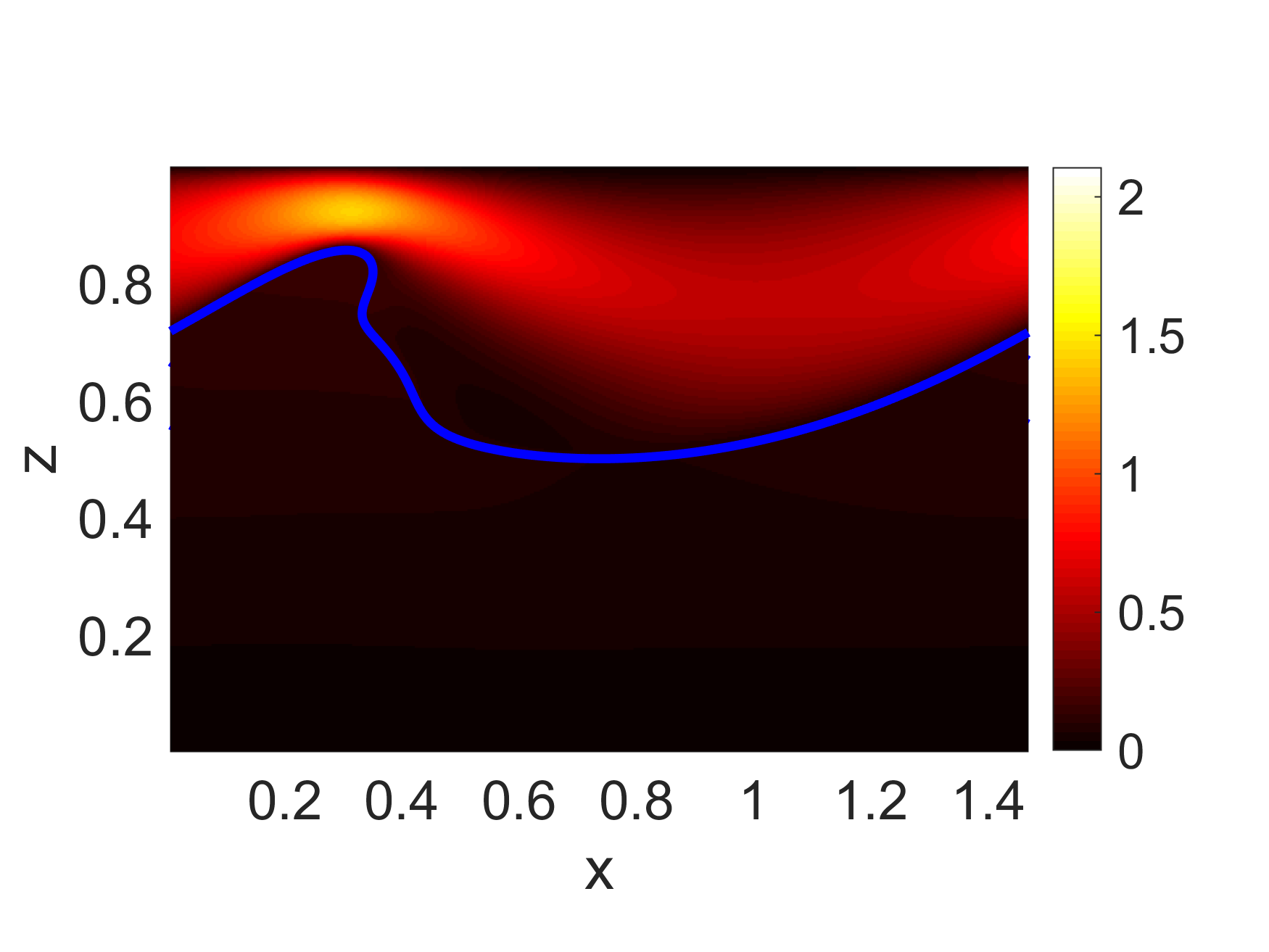}}
		\subfigure[$\,\,\tau=25.2$]{\includegraphics[width=0.45\textwidth]{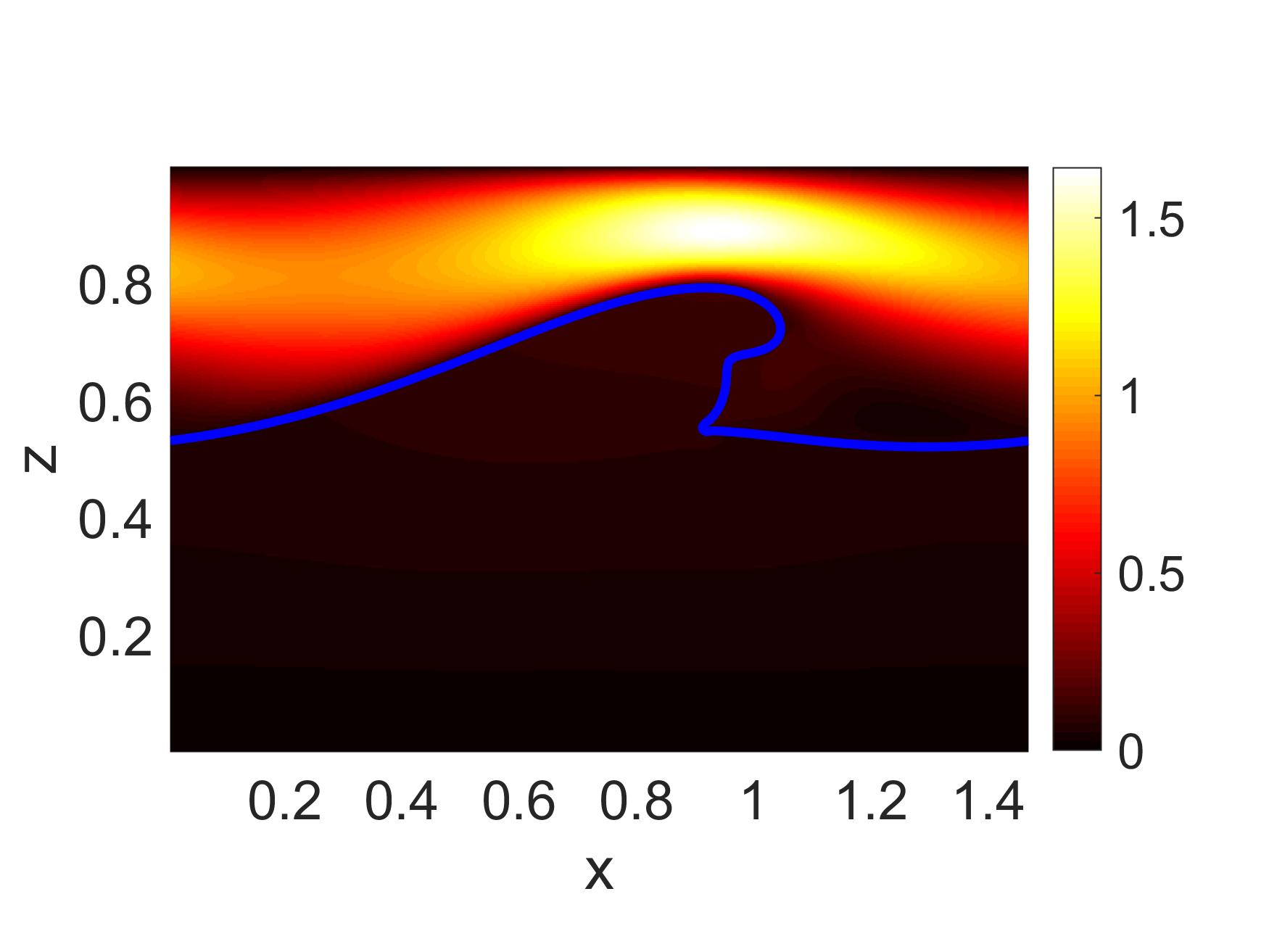}}
		\caption{Snapshots of the interface evolution at different times, obtained via the TPLS numerical solver.  The colorplot shows the $u$-velocity.}
	\label{fig:snapshots_tpls}
\end{figure}

Having confirmed the close agreement between the theory and the TPLS numerical solver for the development of the small-amplitude waves, we continue the simulation into the regime of non-linear waves to the point of wave overturning.  As such, snapshots of the wave evolution are shown in Figure~\ref{fig:snapshots_tpls}.
The multiple harmonic modes present in the initial condition (cf. Equation~\eqref{eq:ic}) degenerate into a single large-amplitude monochromatic wave as the most-dangerous mode asserts itself (Figure~\ref{fig:snapshots_tpls}(a)).  The single remaining wave then steepens and breaks, to form the complicated structure in Figure~\ref{fig:snapshots_tpls}(d)--(e).  A cusp forms at the foot of the wave crest in Figure~\ref{fig:snapshots_tpls}(e).  The cusp gives rise to a large capillary pressure at the cusp point, which makes the simulation highly numerically unstable.  We therefore stop the simulation at the onset of the cusp.

In order to explore what happens beyond the onset of the cusp in Figure~\ref{fig:snapshots_tpls}(e) we resort to a complementary numerical method (ANSYS Fluent; with the numerical setup as described in Section~\ref{sec:method}).  The simulations we consider use $93,697$ computational cells -- this is comparable to the number of grid-points used in the TPLS simulation.    This is sufficient for numerical convergence (a grid-refinement study is presented in Appendix~\ref{app:cvg}).   The time-stepping method is implicit: we use a time-step $\Delta \tau=0.005$ (non-dimensional time units), with typically 20 iterations per time-step required for convergence of the implicit time-stepping method.

\begin{figure}
	\centering
		\subfigure[$\,\,\tau=0$]{\includegraphics[width=0.32\textwidth]{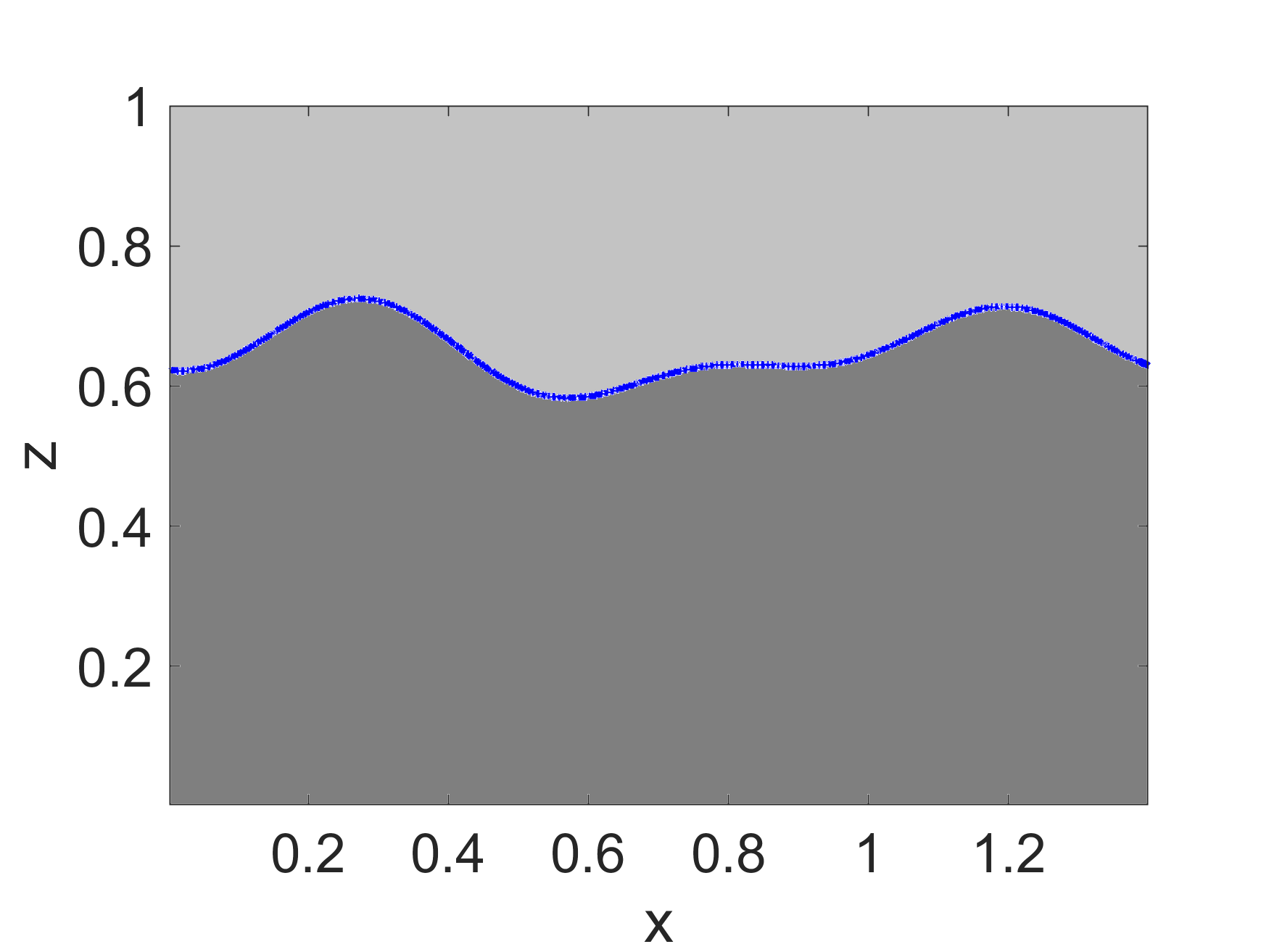}}
		\subfigure[$\,\,\tau=17.5$]{\includegraphics[width=0.32\textwidth]{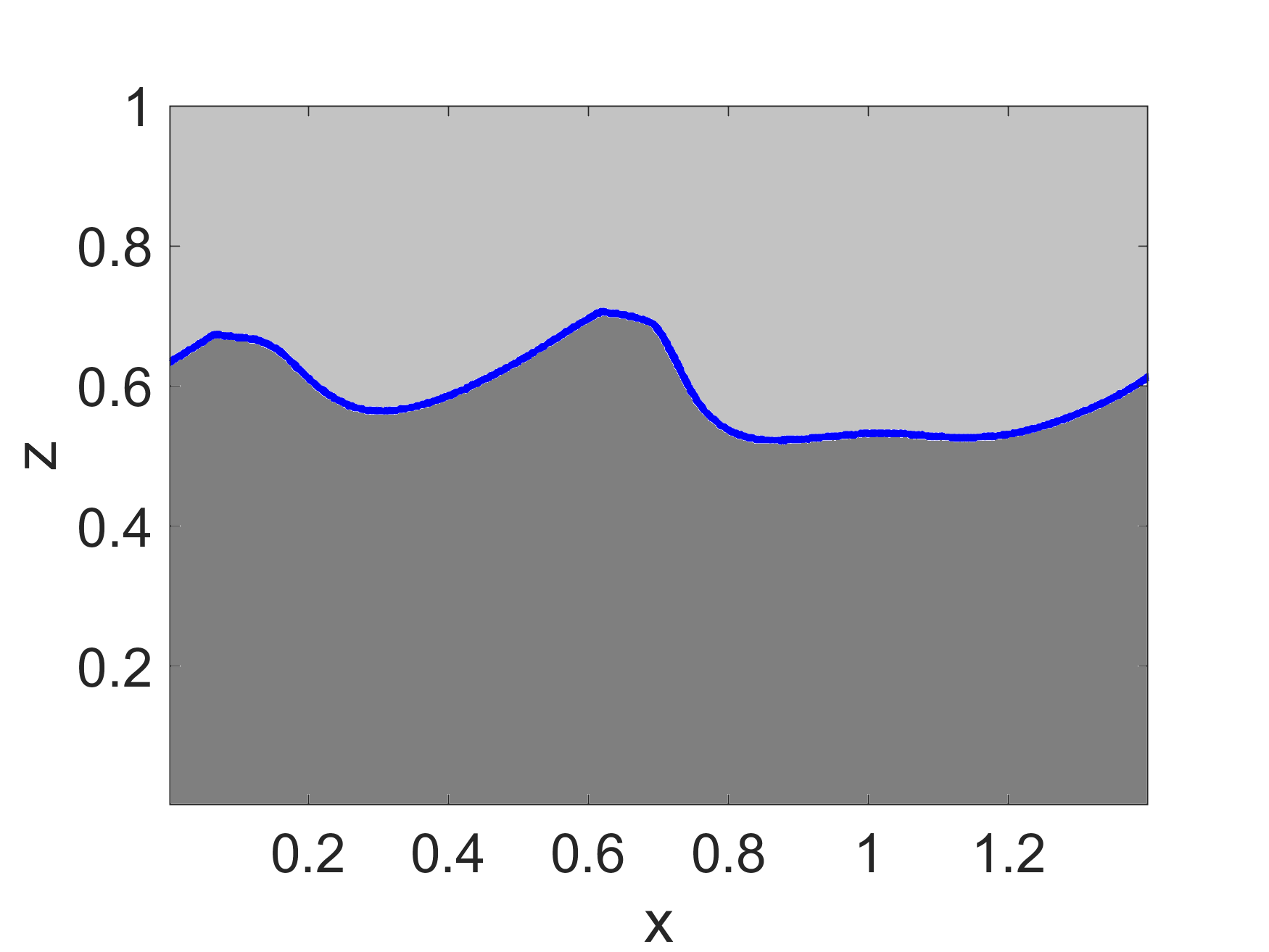}}
		\subfigure[$\,\,\tau=24$]{\includegraphics[width=0.32\textwidth]{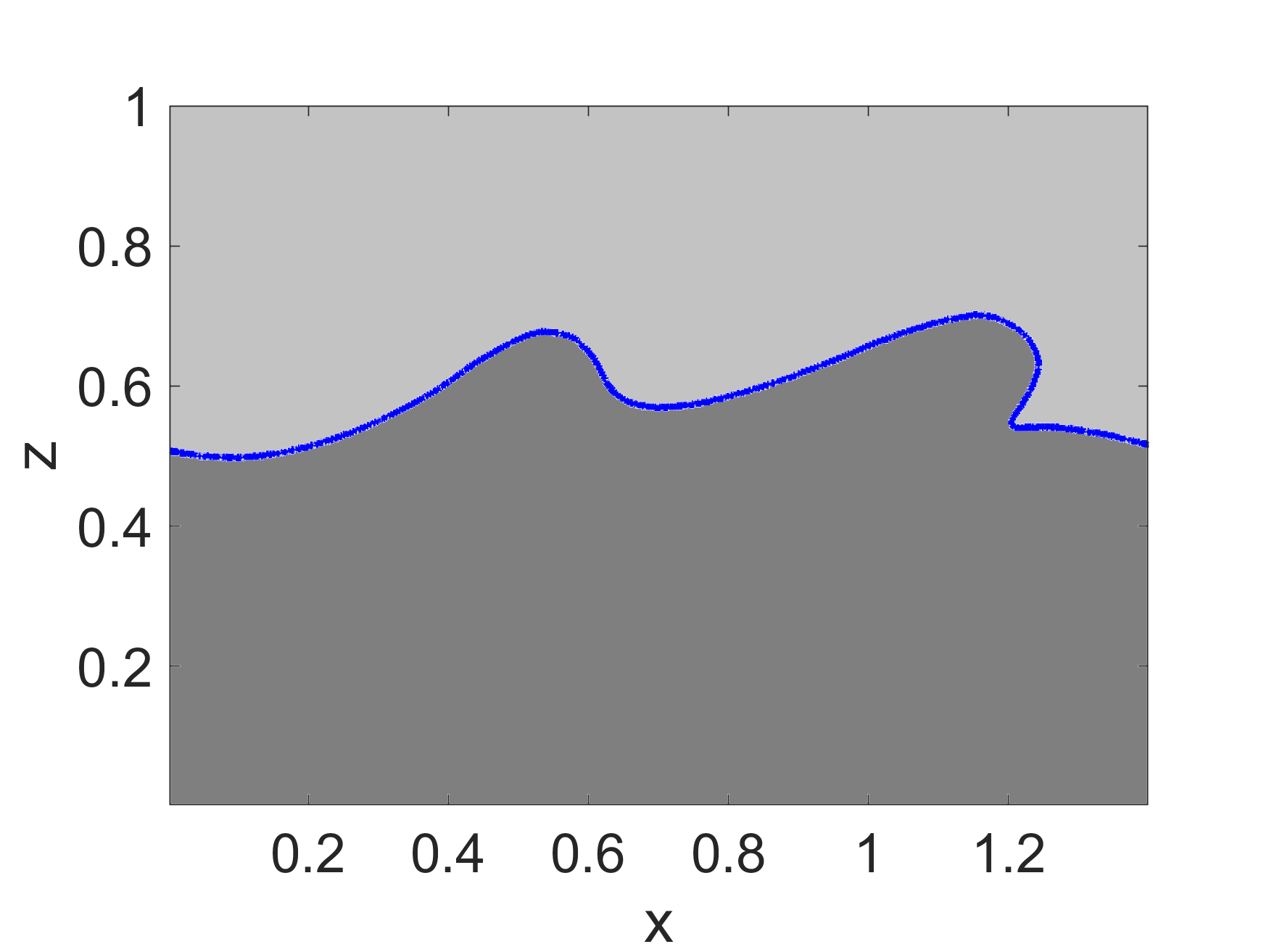}}
		\subfigure[$\,\,\tau=30$]{\includegraphics[width=0.32\textwidth]{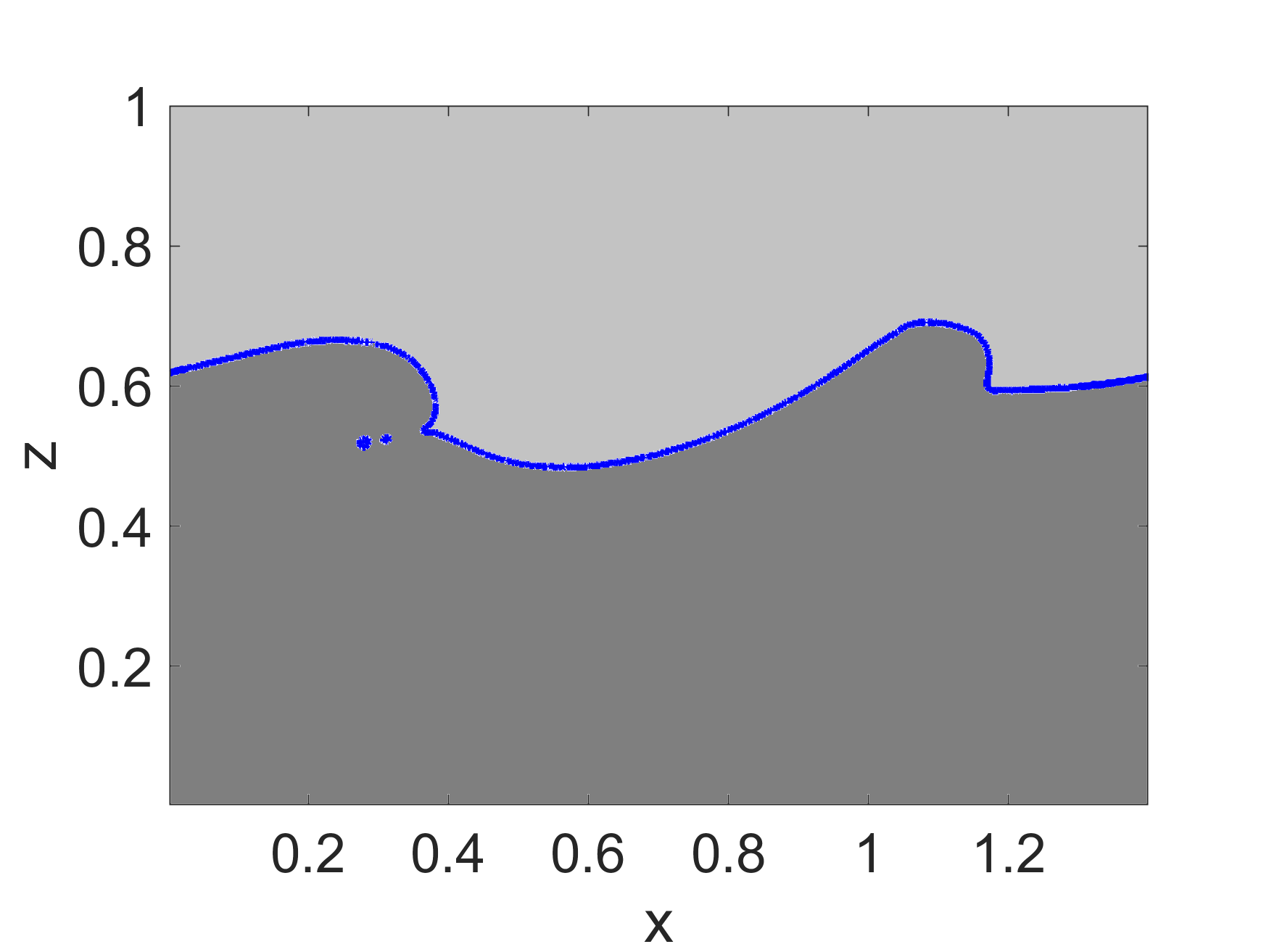}}
		\subfigure[$\,\,\tau=36.4$]{\includegraphics[width=0.32\textwidth]{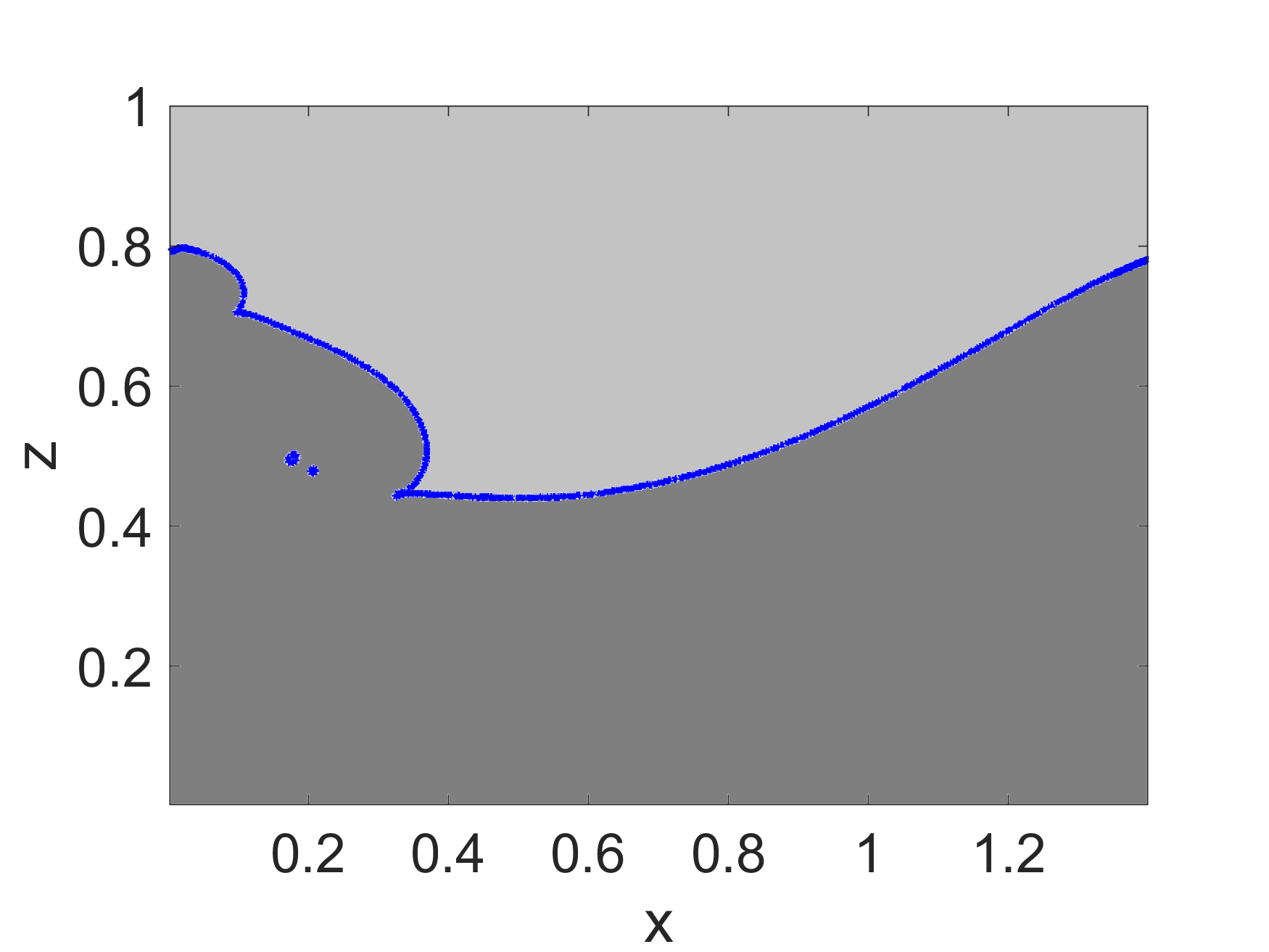}}
		\caption{Snapshots of the interface evolution at different times, obtained via ANSYS Fluent.  Dimensionless quantities are used in the simulation, meaning this figure may be compared directly with Figure~\ref{fig:snapshots_tpls}.  There is no colour bar: the different colours in the snapshots are included just to guide the eye and demarcate the phases.}
	\label{fig:snapshots_ansys}
\end{figure}

We again use the initial condition~\eqref{eq:ic} for the fluid interface (the initial velocities are again zero and the initial pressure corresponds to a simple pressure drop in the $x$-direction).  We take $\epsilon/H=0.1/3$ and $N=3$ -- this `trips' the simulation into a nonlinear state from the very beginning, thereby speeding up the computation.  Snapshots of the interface profile generated with ANSYS Fluent are shown in Figure~\ref{fig:snapshots_ansys}.
The use of the different numerical and analytical methodologies is complementary and inspires confidence in our results:
\begin{itemize}[noitemsep]
\item Starting with the analytical numerical method, this is valid rigorously for the small-amplitude  waves; the implementation of this theory is very well established in the literature~\cite{Boomkamp1996, Onaraigh2013b}.
\item The TPLS numerical simulations agree with the analytical numerical methods for the small-amplitude initial disturbances.  This supports our use of TPLS at early simulation times, up to the formation of the cusp point in Figure~\ref{fig:snapshots_tpls}(d).
\item The ANSYS Fluent simulations agree qualitatively with the TPLS results up to and including the formation of cusps, thereby inspiring confidence in this approach also.  
\end{itemize}
The agreement between the  TPLS simulations and the ANSYS simulations may be inferred by  comparing Figure~\ref{fig:snapshots_tpls} (TPLS) and Figure~\ref{fig:snapshots_ansys}(a--d) (ANSYS).  In both sets of simulations, we have used the same dimensionless variables, hence a comparison between the two simulations is feasible.  The behaviour of the interface in both sets of simulations is qualitatively the same: this suggests that the two approaches are mutually consistent.  Notice however that a quantitative comparison is not possible: a snapshot of a TPLS simulation result at a particular time time $\tau$ may not agree with a snapshot of an ANSYS simulation at the same time, since both sets of simulations use slightly different initial conditions.
Crucially, the simulations in ANSYS may be continued beyond the point of wave overturning (e.g. Figure~\ref{fig:snapshots_ansys}(e,f) -- at these later times, the breaking wave is simply drawn back towards the interface under the influence of gravity, and a final state is a complicated wavy interface -- but no ligament formation.  

\subsection*{Discussion}

It is of interest to compare the results in Figures~\ref{fig:snapshots_tpls}--\ref{fig:snapshots_ansys} with the experimental results of Hu and 
Cubaud~\cite{HuCubaud2018}, who observed ligament formation (such as that shown schematically in Figure~\ref{fig:sketch1} herein) at exactly the same the flow-rates as the ones used in this section.  Therefore, the current numerical and theoretical model under-predicts the experimentally-observed instability.
The origin of the under-prediction can be traced back to the geometry used in the experiments: this is markedly different from the geometry of the present numerical and theoretical model: the experimental microchannel in Reference~\cite{HuCubaud2018} has bounding walls in both the $z$-direction, and the $y$-direction (For the geometric conventions assumed herein, see Figure~\ref{fig:sketch1}).  In contrast, our numerical and theoretical model essentially has periodic boundary conditions in the $y$-direction. 
The relationship between the layer depth the the flow-rate ratio $\varphi=Q_1/Q_2$ is dramatically different depending on the these factors; specifically, this can be seen in Figure~\ref{fig:compare_fig1b_for_paper}.
\begin{figure}
	\centering
		\includegraphics[width=0.6\textwidth]{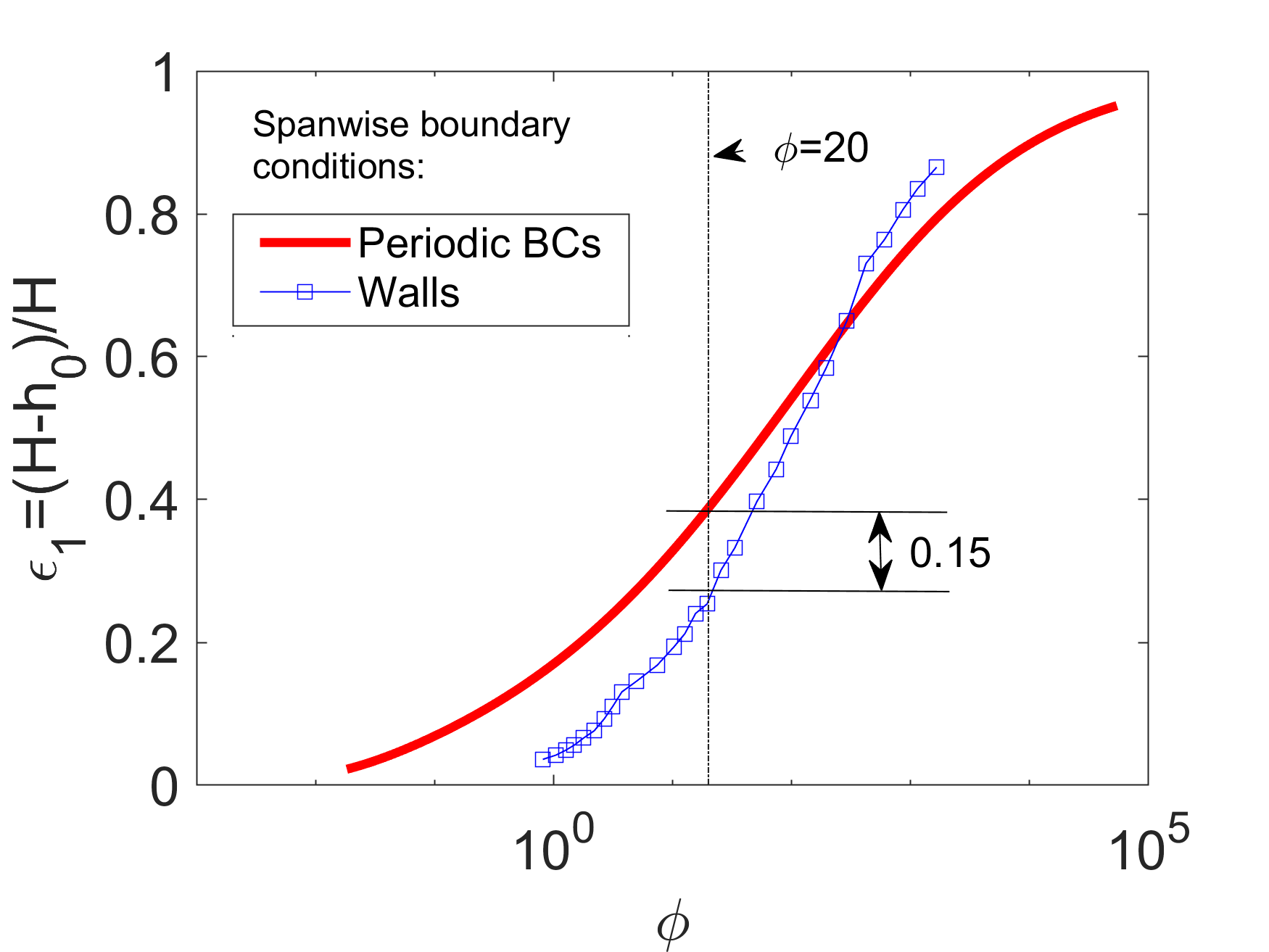}
		\caption{Plot of non-dimensional upper-layer depth $\epsilon_1=(H-h_0)/H$ as a function of the ratio of flow rates $\varphi=Q_1/Q_2$ (solid line).  The value $\varphi=20$ is highlighted, corresponding to the special case {\textbf{Circ}} considered in this section.  A comparison is given with the corresponding functional form for the wall geometry (squares).}
	\label{fig:compare_fig1b_for_paper}
\end{figure}
Since the stability proprieties of the fluid depend not only on the flow rates, but also, independently on the layer depths, it is not surprising that a discrepancy emerges between the current two-dimensional model and the three-dimensional experimental results.  Indeed, previous work on Linear Stability Analysis~\cite{Valluri2010} indicates that increasing $\epsilon_1$ is destabilizing.  Therefore, Figure~\ref{fig:compare_fig1b_for_paper} suggests that the ``wall'' geometry should be more intrinsically unstable than the ``periodic'' geometry.  This is consistent with the contrasting observations in the present work and the experimental observations in of Hu and Cubaud.

\begin{figure}
	\centering
\includegraphics[width=0.6\textwidth]{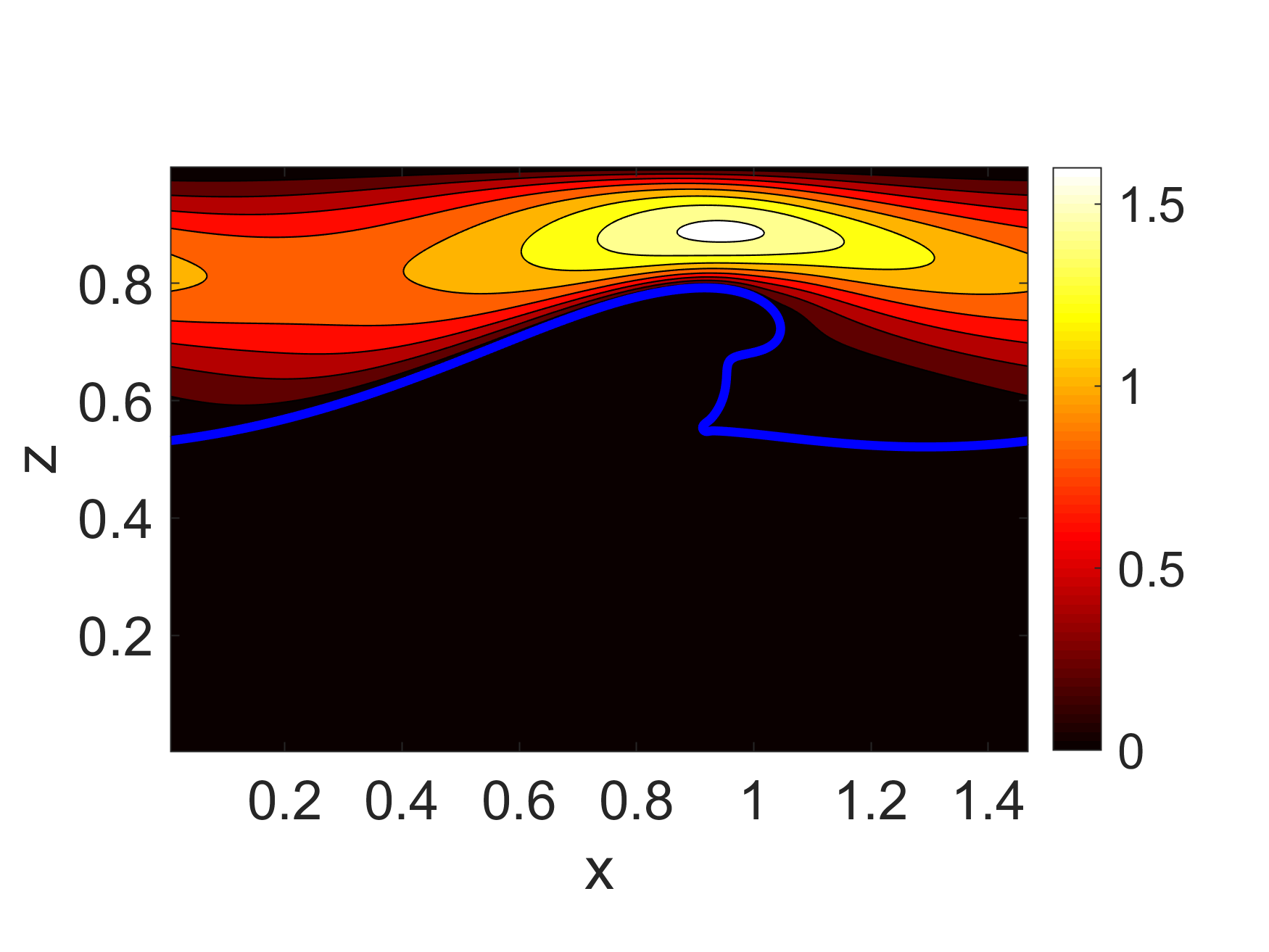}
\caption{Streamlines of the total velocity field at $\tau=25.2$, to be looked at in conjunction with Figure~\ref{fig:snapshots_tpls}.}
	\label{fig:plot2517000_uu}
\end{figure}

As a final word it can be emphasized that even in the current model geometry geometry (which under-predicts the observed instability) produces a strong recirculation flow in the upper layer (e.g. Figure~\ref{fig:plot2517000_uu}) -- this may be of use in microfluidic applications which require either heat transfer, or mixing.

\section{Conclusions}
\label{sec:conc}

Summarizing, we have outlined how analytical and numerical modeling to describe parallel viscous two-phase flows in microchannels.  The focus has been on idealized two-dimensional geometries, with a view to validating the various methodologies for future work in three dimensions.  In the first instance, we have used analytical Orr--Sommerfeld theory to describe the linear instability which governs the formation of small-amplitude waves in such systems.  We have carefully constructed a series of flow-pattern maps to characterize the unstable interfacial waves as a function of the flow rates of the two phases.

We have compared the reslts of the linear stability analysis with the numerical simulations from TPLS; excellent agreement is obtained.   However, the simulations from TPLS are valid well beyond the limit of applicability of Orr--Sommerfeld theory.    We have therefore  continued the numerical simulations into the regime of finite-amplitude interfacial waves, in this way we have exhibited the phenomenon of reverse entrainment whereby droplets of the upper phase are entrained into the lower phase.    We justify our simulations further by comparing the numerical results with corresponding results from a commercial CFD code.  This comparison is again extremely favourable.

In view of the idealized two-dimensional geometry in the present study, a direct comparison with experiments is not possible.  However, experiments in microchannels of a similar size do reveal interfacial waves, as well as wave overturning, ligament formation, and droplet entrainment.  The rigorous validation of the various numerical-simulation techniques in this work pave the way for extending the simulations in future to more realistic geometries, thereby  making a direct comparison with experiments more feasible.

\subsection*{Acknowledgements}

This work has been produced as part of ongoing work within the ThermaSMART network.  The ThermaSMART network has received funding from the European Union's Horizon 2020 research and innovation programme under the Marie Sklodowska--Curie grant agreement No. 778104.  The authors also acknowledge the DJEI/DES/SFI/HEA Irish Centre for High-End Computing (ICHEC) for the provision of computational facilities and support

\appendix
\section{Full formulation of the linear stability analysis}
\label{app:lsa}

In this Appendix, we give a detailed formulation of the governing equations underlying the linear stability analysis in Section~\ref{sec:method} of the main part of the paper.  The starting-point is the base-state (Section~\ref{sec:base}), characterized by a flat interface $z=h_0$ and a laminar flow in the streamwise direction in each phase, denoted by $U_{0i}$.  The flow in the base state is perturbed by the presence of a small-amplitude sinusoidal wave at the otherwise flat interface, such that the location of the perturbed interface reads:
\begin{equation}
z=h_0+\eta(x,t),\qquad \eta(x,t)=\eta_0\mathe^{\imag \alpha x+\lambda t}
\label{eq:app:pert1}
\end{equation}
where $\eta_0$ is a small complex-valued amplitude with $|\eta_0|\ll h_0$, $\alpha$ is the streamwise wavenumber, and $\lambda$ is the growth rate of the disturbance.  The perturbation in Equation~\eqref{eq:app:pert1} gives rise to a perturbation in the velocity and pressure fields -- the complex-valued constant $\eta_0$ in Equation~\eqref{eq:app:pert1} allows for a non-trivial phase relationship between the perturbation velocity and the perturbed interface height, which is determined by the following analysis.

  For reasons alluded to in Section~\ref{sec:method}, it suffices to look at two-dimnesional perturbations characterized by a single wavenumber $\alpha$ in the streamwise direction.  Hence, the perturbed flow can be described by a streamfunction $\psi_i(x,z,t)$, such that
\begin{equation}
\psi_i(x,z,t)=\mathe^{\imag \alpha x+\lambda t}\Psi_i(z),
\label{eq:app:pert2}
\end{equation}
and such that the perturbation velocities $\delta u_i$ and $\delta w_i$ in each phase read:
\begin{equation}
\delta u_i=\frac{\partial\psi_i}{\partial z},\qquad \delta w_i=-\frac{\partial \psi_i}{\partial x}.
\label{eq:app:pert3}
\end{equation}
Here, $\delta u_i$ denotes the perturbed streamwise velocity and $\delta w_i$ denotes the perturbed wall-normal velocity

The equations~\eqref{eq:app:pert1}--\eqref{eq:app:pert3} are substituted into the linearized Navier--Stokes equations (linearized around the base state in Section~\ref{sec:base}).  In this way, we obtain the following set of governing equations:
\begin{subequations}
\begin{eqnarray}
\imag\alpha \rho_2 \left[\left(\Psi_2''-\alpha^2\Psi_2\right)\left(U_{02}-c\right)-\Psi_2 U_{02}''\right]&=&\mu_2\left(\Psi_2''''-2\alpha^2\Psi_2''+\alpha^4\Psi_2\right),\\
\imag\alpha \rho_1 \left[\left(\Psi_1''-\alpha^2\Psi_1\right)\left(U_{01}-c\right)-\Psi_1 U_{01}''\right]&=&\mu_1\left(\Psi_1''''-2\alpha^2\Psi_1''+\alpha^4\Psi_1\right),
\end{eqnarray}%
\label{eq:app:osbulk}%
\end{subequations}%
Here, the growth rate $\lambda$ has been rewritten in terms of the complex wave speed $c$, via the identity
\begin{equation}
\lambda=-\imag \alpha c.
\label{eq:app:lambdadef}
\end{equation}
Equations~\eqref{eq:app:osbulk} are supplemented with the following no-slip and no-penetration boundary conditions:
\begin{equation}
\Psi_i=\Psi_i'=0,
\label{eq:app:bcs}
\end{equation}
at the walls $z=0$ and $z=1$.  

In addition,  matching conditions are prescribed at the interface $z=h_0+\eta$, with $\eta=\eta_0\mathe^{\imag \alpha x+\lambda t}$.  In the streamwise direction, continuity of velocity and tangential stress  (\textit{cf.} Equation~\eqref{eq:jump}) imply the following relations: 
\begin{subequations}
\begin{eqnarray}
\Psi_2&=&\Psi_2,\\
\Psi_2'+\eta U_{02}'&=&\Psi_1'+\eta U_{01}',\\
\mu_2\left(\Psi_2''+\alpha^2\Psi_2\right)&=&\mu_1\left(\Psi_1''+\alpha^2\Psi_1\right).
\end{eqnarray}
The perturbed interface location can be determined from the kinematic condition, which requires that the interface moves with the flow: $\partial_t \eta+U_{0}\partial_x\eta=w$ (the subscripts are suppressed because both the velocities in each phase are the same at the interface).  In terms of streamfunctions, the kinematic condition reads:
\begin{equation}
\eta_0=\Psi_2/(c-U_{02})=\Psi_1/(c-U_{01});
\end{equation}
this determines the phase relationship between $\eta=\eta_0\mathe^{\imag \alpha x+\lambda t}$ and the perturbed velocity fields.

The remaining interfacial matching condition arises from imposing a linearized jump condition on the normal stress at the interface (\textit{cf.} Equation~\eqref{eq:jump}):
\begin{multline}
\imag \alpha \rho_2 \left[\Psi_2\left(c-U_{02}\right)+\Psi_2 U_{02}'\right]+\mu_2\left(\Psi_2'''-3\alpha^2\Psi_2\right)\\
=\imag \alpha \rho_1 \left[\Psi_1'\left(c-U_{01}\right)+\Psi_1 U_{01}'\right]+\mu_1\left(\Psi_1'''-3\alpha^2\Psi_1\right)
%
%=\frac{ k^4}{\mywe}
+\alpha^2\left(g+\gamma\alpha^2\right)
\left[\frac{\Psi_1'-\Psi_2'}{\imag\alpha\left(U_{02}'-U_{01}'\right)}\right]=0.
\end{multline}%
\label{eq:app:imcs}%
\end{subequations}%

Equations~\eqref{eq:app:osbulk}--\eqref{eq:app:imcs} constitute an eigenvalue problem for the streamfunction components $(\Psi_2,\Psi_1)$, with eigenvalue $\lambda=-\imag\alpha c=-\imag\omega$.  They can be formulated in an operator/matrix form given and hence, solved  numerically using standard Chebyshev collocation techniques~\citep{Boomkamp1997}.  This method has been further developed and validated in the context of viscous liquid-liquid flows in Reference~\cite{Naraigh2014linear} and is therefore used in the main part of the paper without further commentary.

\subsection*{Energy-budget analysis}

To understand the physical mechanism that causes the instability, we perform an energy-budget analysis.  We multiply the corresponding linearized equations of motion for the perturbation velocity $\delta\vecu=(\delta u,\delta w)$ by $\delta\vecu$ itself and integrate over the $x$- and $z$-directions (the corresponding perturbation pressure is denoted by $\delta p$).  The $x$-variable is integrated over a single wavelength $\myell=2\pi/\alpha$ and the $z$-variable is integrated over the full vertical extent of the channel.  In a standard fashion, this gives the following energy-budget relation
\begin{equation}
\mathcal{P}=	\sum_{i=1}^2{KIN}_i=\sum_{i=1}^2{REY}_i+\sum_{i=1}^2{DISS}_i+{INT},
\end{equation}
%
%where
%
\begin{eqnarray*}
{KIN}_i&=&\tfrac{1}{2}\frac{\mathd}{\mathd t}\iint \rho_j|\delta\vecu_j|^2\,\mathd x\mathd z,\\
{REY}_i&=&-\rho_i \iint \delta u_i\delta w_iU_{0i}'\,\mathd x\mathd z,\\
{DISS}_i&=&-\mu_i\iint \left[
2\left(\frac{\partial }{\partial x}\delta u_i\right)^2
+2\left(\frac{\partial }{\partial z}\delta w_i\right)^2
+\left(\frac{\partial }{\partial z}\delta i+\frac{\partial }{\partial
x}\delta w_i\right)^2\right]\,\mathd x\mathd z.
\end{eqnarray*}
%
%Here, $j=T,B$, with $r_T=1$ and $r_B=r$, and similarly for $m_j$.
The term ``${INT}$'' is related to interfacial conditions, and is given in terms of the following stress tensor for the perturbed flow:
\begin{equation}
	\Ttensor_{xx,i}=-\delta p_i+2\mu_i \frac{\partial }{\partial x}\delta u_i,\qquad
	\Ttensor_{zz,i}=-\delta p_i+2\mu_i \frac{\partial }{\partial z}\delta w_i,\qquad
	\Ttensor_{xz,i}=\mu_i\left(\frac{\partial }{\partial z}\delta u_i+\frac{\partial}{\partial x}\delta w_i\right).
\end{equation}
Thus,
\begin{equation}
	{INT} =
	\int_0^\myell\left[\delta u_2 \Ttensor_{xz,2}+\delta w_2\Ttensor_{zz,2}\right]_{z=0}\mathd x
	-\int_0^\myell\left[\delta u_1 \Ttensor_{xz,1}+\delta w_1\Ttensor_{zz,1}\right]_{z=0}\mathd x,
\end{equation}
which is decomposed into normal and tangential contributions,
\begin{equation}
	{INT}={NOR}+{TAN},
\end{equation}
where
\begin{equation}
	{NOR}=\int_0^{\myell}\left[\delta w_2\Ttensor_{zz,2}-\delta w_1\Ttensor_{zz,1}\right]_{z=0}\mathd x,
\end{equation}
and
\begin{equation*}
	{TAN}=\int_0^{\myell}\left[\delta u_2\Ttensor_{xz,2}-\delta u_1 \Ttensor_{xz,1}\right]_{z=0}\mathd x.
\end{equation*}

\section{Sample convergence study}
\label{app:cvg}

\begin{figure}
	\centering
		\subfigure[$\,\,\tau=0$]{\includegraphics[width=0.32\textwidth]{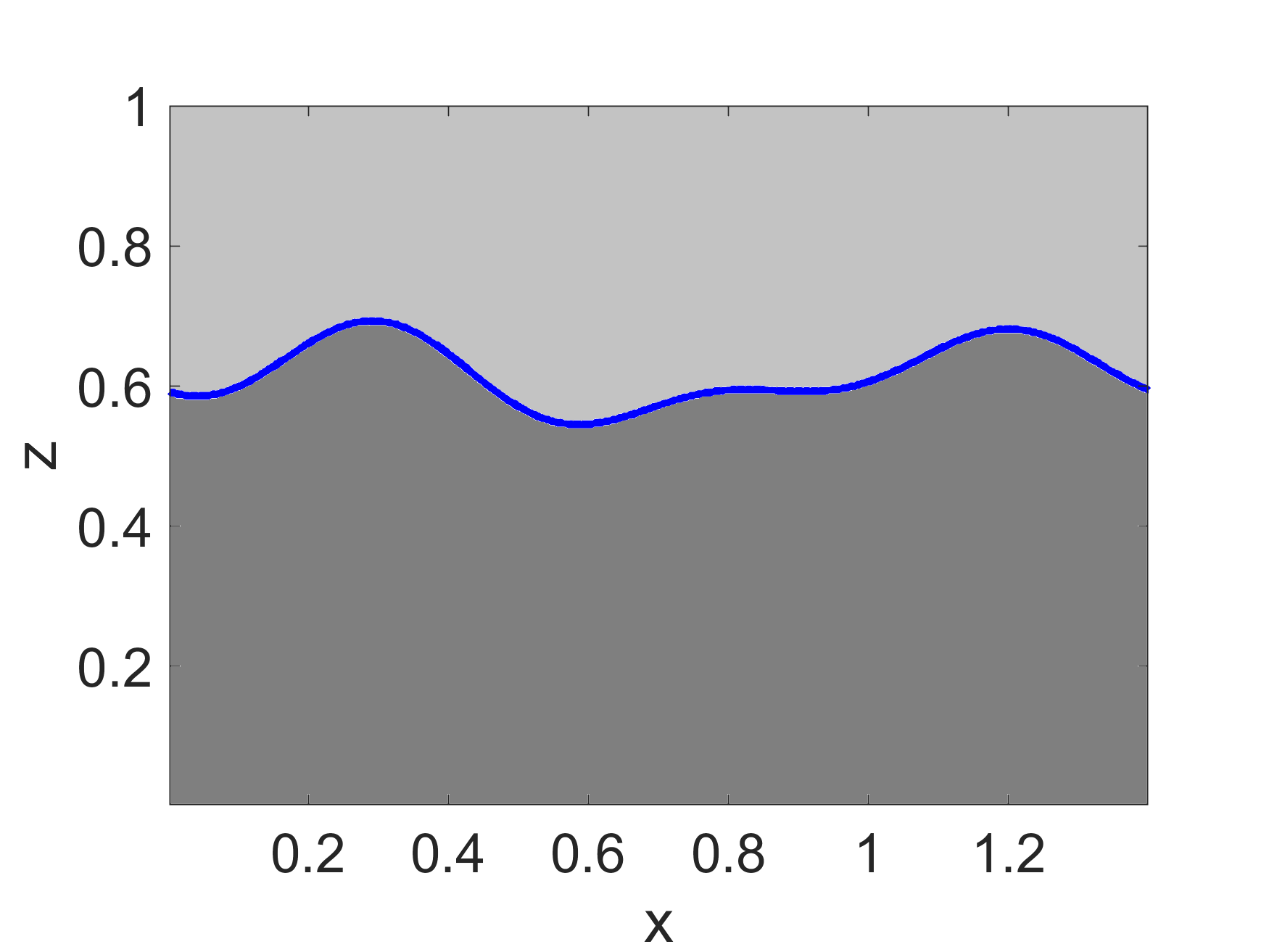}}
		\subfigure[$\,\,\tau=6.25$]{\includegraphics[width=0.32\textwidth]{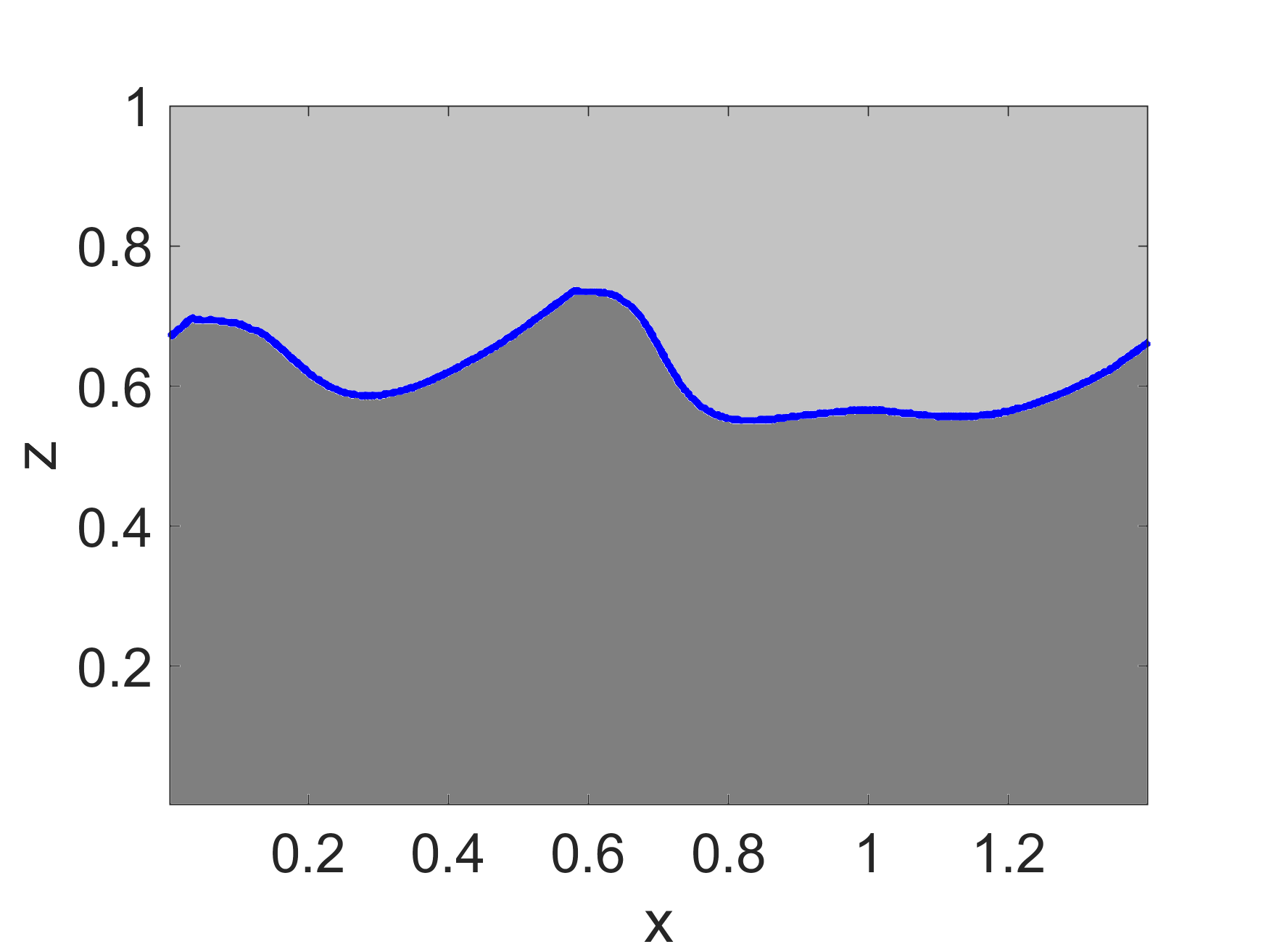}}
		\subfigure[$\,\,\tau=11.25$]{\includegraphics[width=0.32\textwidth]{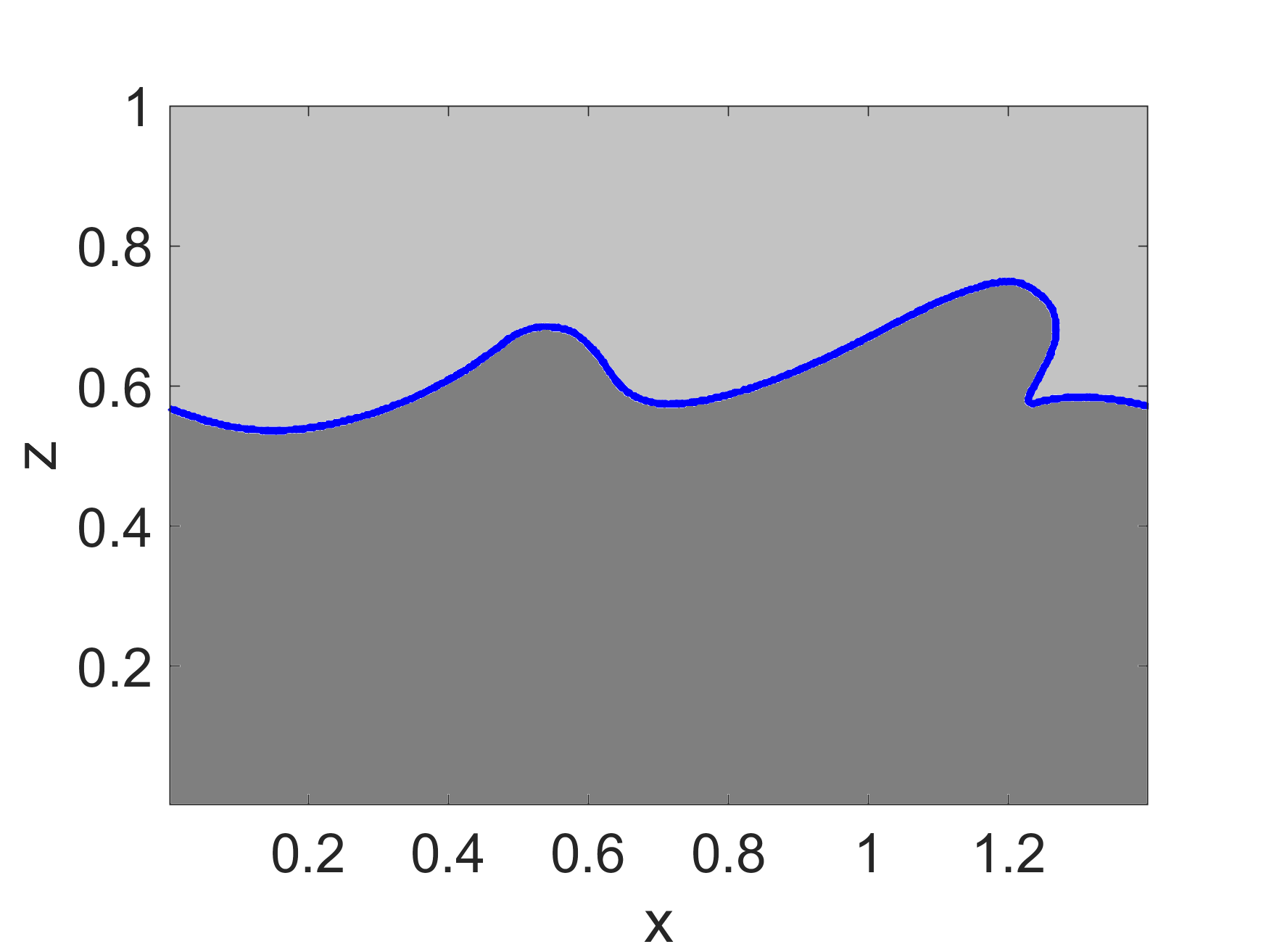}}
		\subfigure[$\,\,\tau=16.25$]{\includegraphics[width=0.32\textwidth]{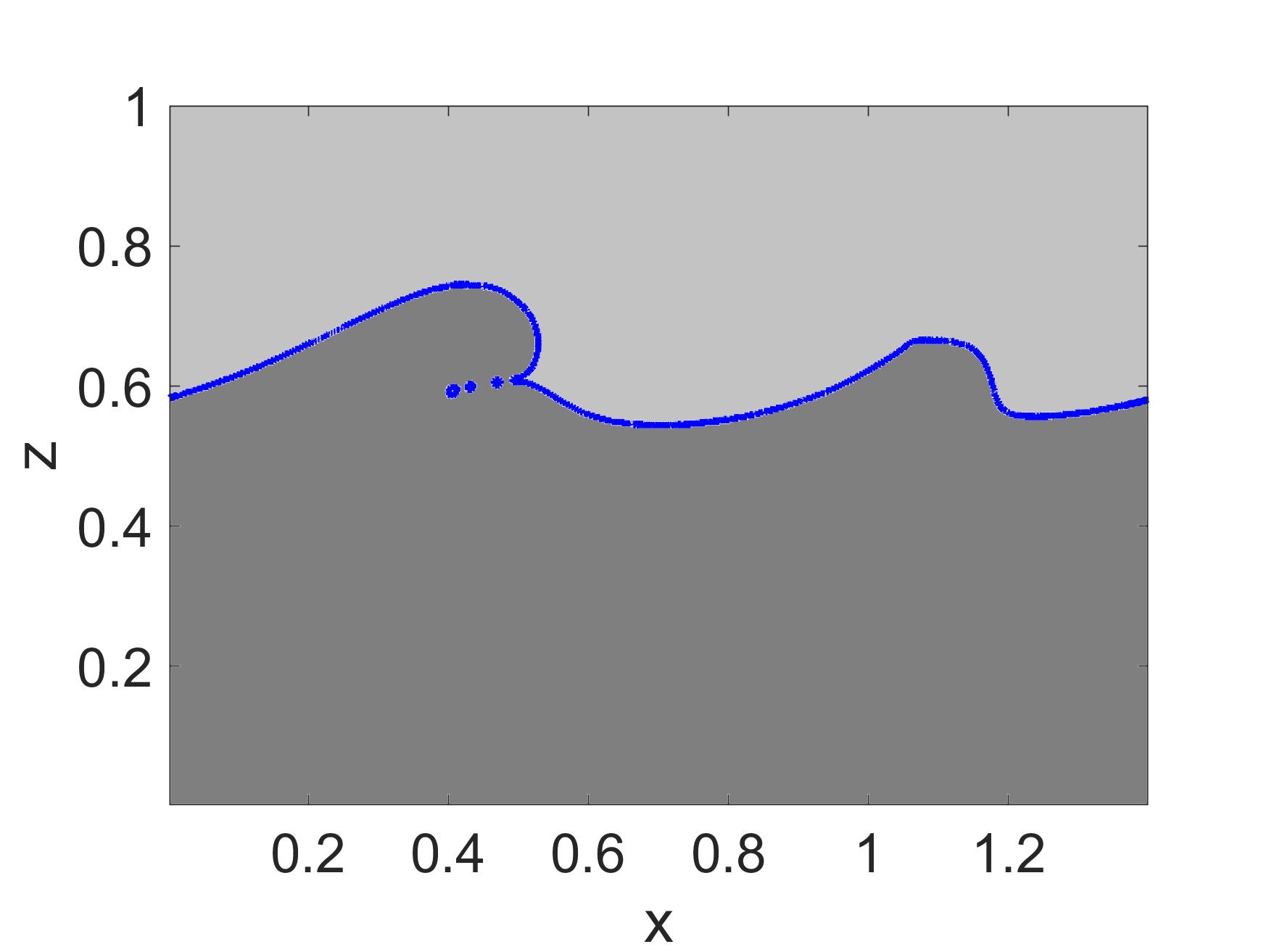}}
		\subfigure[$\,\,\tau=33.75$]{\includegraphics[width=0.32\textwidth]{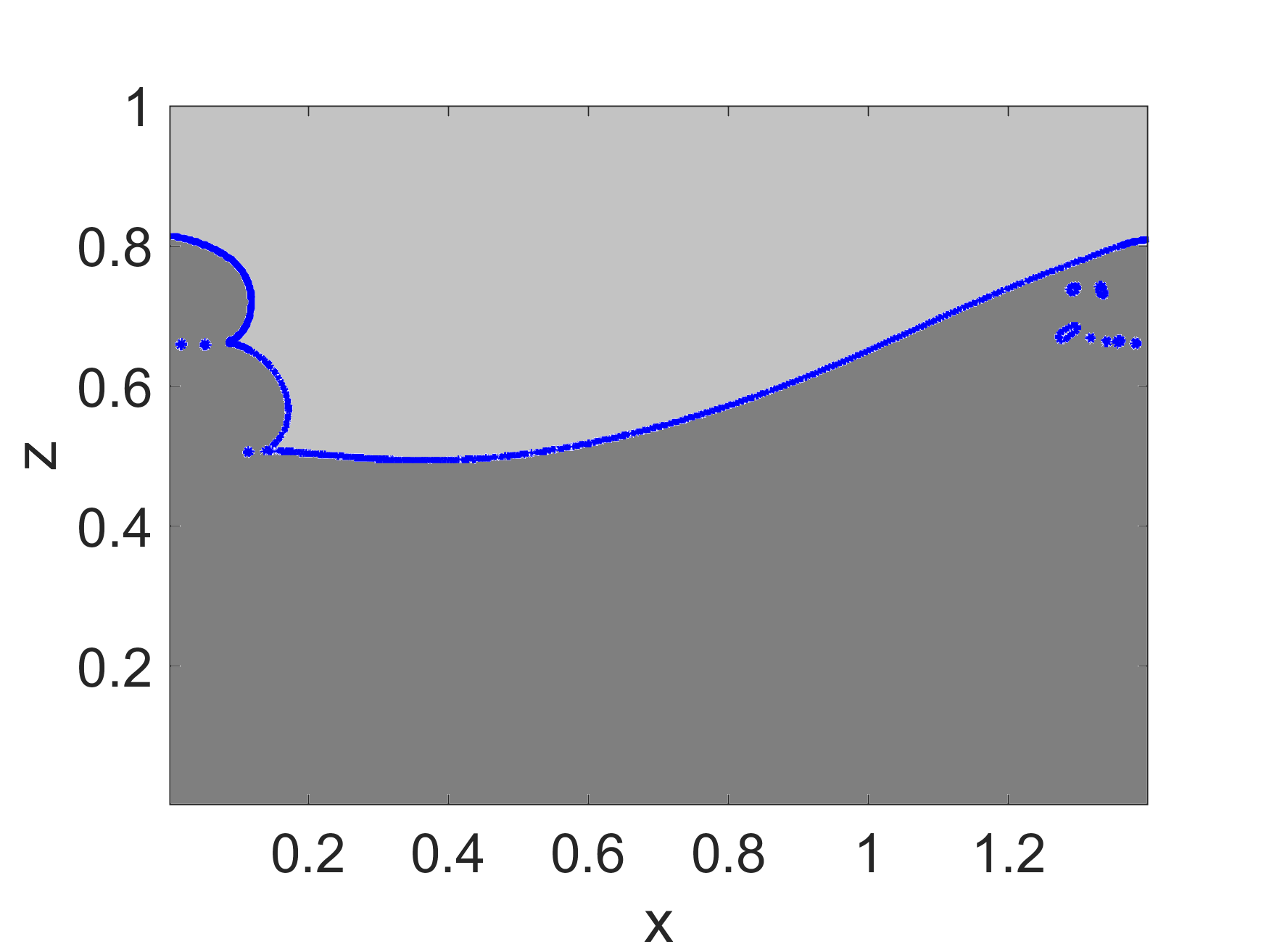}}
		\caption{Snapshots of the interface evolution at different times, obtained via ANSYS Fluent -- high-resolution simulation, 178,867 computational cells.   The figure may be compared directly with the corresponding low-resolution simulation in Figure~\ref{fig:snapshots_ansys}.  }
	\label{fig:snapshots_ansys_x}
\end{figure}

In order to validate the convergence of our numerical results, we have carried out a grid-refinement study on the case study involving the ANSYS simulations.  As such, we have re-run the simulations corresponding to Figure~\ref{fig:snapshots_ansys} ($93,697$ computational cells) at a much higher resolution (178,867 computational cells).  The results of the high-resolution simulation are shown in Figure~\ref{fig:snapshots_ansys_x}.  The snapshots in these two figures may be compared panel-by-panel.
Excellent agreement between the two figures can be seen, with the exception of panels (b) in the figures, where the mismatch is due to the different ways in which the two simulations have been initialized.
The main point is that the two sets of results are almost identical.
This confirms the robustness of the presented simulation results to grid refinement.  In particular, it can be noted that the reverse-entrainment effect is visible in both Figures~\ref{fig:snapshots_ansys}(f) and~\ref{fig:snapshots_ansys_x}(f), confirming that this is a physical effect and not an unphysical effect due to lack of grid resolution.
Furthermore, as the TPLS results presented in the main text are in agreement with both Orr--Sommerfeld Theory and the low-resolution ANSYS simulations, it can be concluded that the TPLS results are independent of the grid resolution.

%\bibliographystyle{unsrt}
%\bibliography{turbulence_bibliography}

\end{document}